\documentclass{aa}  
\usepackage[table]{xcolor}
\usepackage{multicol}
\usepackage{multirow}
\usepackage{booktabs}
\usepackage{graphicx}
\usepackage{txfonts}
\usepackage{comment}
\usepackage{makecell}
\usepackage[normalem]{ulem}
\usepackage[usestackEOL]{stackengine}
\usepackage[colorlinks=true,allcolors=blue]{hyperref}
\usepackage[capitalize]{cleveref}
\begin{document} 
\newcommand\bluesout{\bgroup\markoverwith{\textcolor{blue}{\rule[0.5ex]{2pt}{0.4pt}}}\ULon}
   \title{SLOW IV: Not all that is Close will Merge in the End - \\Superclusters and their Lagrangian collapse regions}

   \subtitle{}

   \author{B. A. Seidel\inst{1},
           K. Dolag\inst{1,2},
           R.-S. Remus\inst{1},
           J. G. Sorce\inst{3,4,5},
           E. Hernández-Martínez\inst{1},
           I. Khabibullin\inst{1,2},
           N. Aghanim\inst{4}
          }
   \authorrunning{Seidel et al.}

   \institute{Universit\"ats-Sternwarte, Fakult\"at f\"ur  Physik, Ludwig-Maximilians Universität, Scheinerstr. 1, 81679 M\"unchen, Germany\\
   \email{bseidel@usm.uni-muenchen.de}
         \and
         Max-Planck-Institut für Astrophysik, Karl-Schwarzschild-Straße 1, 85741 Garching, Germany
         \and
         Univ. Lille, CNRS, Centrale Lille, UMR 9189 CRIStAL, 59000 Lille, France
          \and
          Université Paris-Saclay, CNRS, Institut d’Astrophysique Spatiale, 91405 Orsay, France\
          \and
          Leibniz-Institut f\"{u}r Astrophysik (AIP), An der Sternwarte 16, 14482 Potsdam, Germany
             }
   \date{Received September 15, 1996; accepted March 16, 1997} 
  \abstract
   {Large scale agglomerations of galaxy clusters are the most massive structures in the universe. To what degree they are actually bound against an accelerating expansion of the background cosmology is of significant cosmological as well as astrophysical interest. In this study, we introduce a cross matched set of superclusters from the SLOW constrained simulations of the local (z<0.05) universe. These simulations combine a central region constrained by local velocity field data and realistic baryonic physics models within a 500 Mpc/h Box to reproduce the locally observed large scale structure in detail.}
   {Identifying the local superclusters provides estimates on the efficacy of the constraints in reproducing the local large-scale structure accurately. The simulated counterparts can help identifying possible future observational targets containing interesting features such as bridges between pre-merging and merging galaxy clusters and collapsing filaments and provide comparisons for current observations. By numerically determining the collapse volumes for the simulated counterparts we further elucidate the dynamics of cluster-cluster interactions in those regions. }
   {Starting from observational catalogs of local superclusters and the most massive clusters from the SLOW simulations already identified in previous works, we search for simulated counterparts of supercluster members of six regions. We evaluate the significance of these detections by comparing the observed geometries to supercluster regions in random simulations. We then run an N-body version of the SLOW initial conditions into the far future and determine which of the member clusters are gravitationally bound to the host superclusters. Furthermore we compute masses and density contrasts for the collapse regions.}
   {We demonstrate the SLOW constrained simulation of the local universe to accurately reproduce local supercluster regions not only in mass of their members but also in the individual clusters three-dimensional geometrical arrangement relative to each other.  We furthermore find the bound regions of the local superclusters consistent in both size and density contrast with previous theoretical studies. This will allow to connect future numerical zoom-in studies of the clusters to the large scale environments and specifically the supercluster environments these local galaxy clusters evolve in. The zoom-ins will focus on ICM properties, turbulence and non-thermal emission and build on the existing work concerned with the environments of local galaxy clusters.}
   {}

   \maketitle
%

\section{Introduction}
Descriptions of the large-scale structure of the universe usually categorize components of the overarching structure, the cosmic web \citep{bond1996a,weygaert2008}, according to their collapse mode. Two dimensional collapsing structures are called sheets, linear collapsing structures are called filaments, and the intersection points of filaments, where the most massive clusters are located are called nodes. 

Superclusters, as arrangements of galaxy clusters, fall somewhat outside of these three categories since their definition is elusive and encompasses a variety of different morphologies, masses and collapse modes. For example, these regions have been dynamically identified with their basins of attraction, regions within which all large-scale flow lines (disregarding the background evolution) converge to a single point \citep{tully2014}. Subsequent work has shown this definition to cleanly partition the entire cosmological volume into discrete regions \citep{dupuy2019} and these regions then encompass matter organized into all three components of the cosmic web and even voids. Other work \citep{einasto2018,einasto2019} defines superclusters as regions enclosed by contours of enhanced density -- or, correspondingly, luminosity -- within the cosmic web itself partitioning nodes and sections of filaments according to their density contrast. Studies of the morphologies of the superclusters obtained with these methods yielded two main morphological groups of superclusters using Minkowski functionals: Spider superclusters formed around one central massive object fed by multiple filaments and filament superclusters where the mass is more evenly distributed along one massive filament giving rise to a more elongated geometry \citep{einasto2007}. \citet{einasto2007} additionally connected these types to the richness of the region, with filaments representing the richest superclusters, while spiders were found to be in general poorer.
An even more restrictive definition is given by the "superstes" cluster, coined by \citet{chon2015}, comprised only of components that are actually gravitationally bound to each other. All of these definitions capture a different aspect of the largest matter arrangements in the universe and can thus yield radically different properties. 

Supercluster regions are of considerable interest from a number of different perspectives: They host a large number of bridges and filaments connecting galaxy clusters in a pre-merging state making them prime targets for a number of thermal and non-thermal emission processes as well as the influence of different environments on galaxy evolution \citep[e.g.][]{aghanim2024}. Additionally these massive structures have been investigated in studies aimed at understanding large scale magnetic fields \citep[e.g.][]{xu2006}. Finally, as the largest coherent density perturbations in the universe, they decouple from the Hubble flow last and in fact only a few of these objects have been shown to be already actively collapsing\citep{pearson2013,pearson2014,reisenegger2000}. This relatively pristine state allows studies of the primordial configuration of matter density perturbations from a cosmological perspective.

The local universe in particular is host to numerous peculiarities that possibly can challenge the current canonical cosmological model but at the very least point towards the local volume as being an outlier in terms of large-scale structure. Examples for such features are the local sheet \citep{einasto1983,bohringer2021} which is a 100~Mpc pancake in the local universe, a significant local over- \citep{makarov2011} or underdensity \citep{bohringer2020} depending on the tracer for the underlying matter distribution used and supercluster structures that strain predictions of the cosmological model for the most massive structures in the universe \citep{sheth2011}. These features can play an important role in resolving or contextualizing current tensions between cosmological parameters obtained from local data and derived from the CMB (assuming the standard cosmological model) such as $H_0$ and $\sigma_8$.

The SLOW set of constrained simulations is the first simulation suite to combine constraints from the local velocity fields, a large study volume and detailed galaxy evolution models, making it the ideal laboratory for studying these local peculiarities. While previous works using SLOW studied the clustering properties of various mass tracers in the simulations \citep{dolag2023a}, crossmatched and compared local galaxy clusters to their simulated counterparts \citep{hernandez-martinez2024a} and provided a first estimate of the expected synchotron emission from the local cosmic web \citep{boss2023}, in this work we introduce the local supercluster regions as reproduced by these simulations.
This paper is structured in the following way: In section 2 we briefly introduce the SLOW simulations and explain the process of identifying and validating supercluster cross-matches therein. In section 3 we present the supercluster regions and members obtained with these methods and compare them to observed properties. We furthermore introduce the Clairvoyant simulations, a set of N-body forward simulations aimed at determining the bound-ness of locally identified supercluster regions and determine for each of the identified supercluster members whether they are bound to their supercluster core. Finally we compare the properties of these bound objects to theoretically obtained properties from spherical collapse theory in ~\ref{sec:future}.
\section{Methods}
\subsection{Constrained simulations}
The goal of constrained simulations is to generate large-scale environments that are not only compatible with random Gaussian initial fluctuations but also match the locally observed structures like galaxy clusters, groups and the cosmic web in our cosmic neighborhood. In the past decade this approach to cosmological simulations has gained significant traction with a number of different methods and constraints being used. The approaches can be roughly divided into two groups: Constrained simulations based on galaxy number density information and those based on galaxy peculiar velocity data. An example for the former approach are the Coruscant simulations \citep{dolag2005c}. Additionally constrained simulations based on Bayesian forward modeling \citep{kitaura2012b,hess2013} such as the Sibelius simulations \citep{sawala2022} take galaxy density as an input, but add dynamical modeling. Using velocity data has the distinct advantage of implementing information about an underlying field that fluctuates on a larger scale than the density field, thus providing better large-scale constraints. Simulations using galaxy velocities to build constraints include the CLONES simulations \citep{sorce2018,sorce2021}, which the HESTIA suite \citep{libeskind2020} and the SLOW set of initial conditions and simulations \citep{dolag2023a} are based on. Previous galaxy cluster and large scale studies using constrained realizations obtained with these methods include, but are not limited to: Analysis of the pressure profiles of a digital Virgo clone accounting for projection effects \citep{lebeau2024c} and the variation of the splashback radius of Virgo when estimated with different tracers \citep{lebeau2024a}. Additionally \citet{malavasi2023} studied the connectivity of a simulated Coma cluster counterpart to the cosmic web and compared simulated properties to the observed Coma cluster. Large-scale studies with constrained simulations investigated the local galaxy distribution \citep{mcalpine2022} and velocity waves in the Local Hubble diagram as a diagnostic tool for cluster potentials and masses \citep{sorce2024} and aimed at constraining the local magnetic field \citep{dolag2005a}. On larger scales yet, a possible impact of local clusters on CMB anomalies via the thermal and kinematic Sunyaev--Zeldovich effect was quantified by \citet{jung2024}.

\subsection{Initial conditions}
The method of obtaining the initial conditions is described in detail in previous works \citep{sorce2018,dolag2023a}, thus we limit the discussion here to a brief summary: From the Cosmicflows-2 catalogue of distance moduli and redshifts \citep{tully2013a} the local velocity field is reconstructed using a Wiener Filter technique of variance minimization \citep{zaroubi1995,zaroubi1999}. A number of bias correction procedures are subsequently applied to the raw data to maximize the accuracy and robustness of the reconstruction  and handle the shot noise and intrinsic uncertainty from the Cosmicflows-2 data \citep{doumler2013, sorce2014,sorce2015,sorce2017,sorce2017a,sorce2018a}. After these methods are applied, the final initial conditions are obtained by applying the constrained realization algorithm introduced by \citet{hoffman1991} to the reconstructed velocities. 

\subsection{Simulation suite: SLOW}
Starting from a set of low-resolution constrained initial conditions, the resolution is increased self-consistently by adding shorter wavelength density fluctuation to the matter power spectrum. This is done using the \textsc{Ginnungagap} software\footnote{https://code.google.com/p/ginnungagap/}. Starting from a $512^3$ dark matter particle grid in a box of 500 Mpc/h sidelength (realization 8 from CLONES), SLOW  covers a number of up- and downscaled resolutions. This paper will focus on the SLOW volume currently combining the most complete subgrid physics (including feedback from active galactic nuclei, AGN) with the highest resolution, termed the $\mathrm{SLOW-FP1536}^3$ run. All simulations are performed using the smoothed-particle hydrodynamics (SPH) code \textsc{OpenGadget3}, which expands on the \textsc{Tree-SPH} code \textsc{Gadget2} \citep{springel2005c}. The $\mathrm{SLOW-FP1536^3}$ run employs a subgrid model largely identical to the  \textit{Magneticum} suite of simulations \citep{dolag2016b}: The baryonic physics modules employ an updated SPH formalism \citep{beck2016} additionally implementing higher-order kernels \citep{dehnen2012}. The physical processes necessary for galaxy evolution, such as cooling, star formation and winds are modeled based on the hybrid multi-phase model \citep{springel2003b}. Extensions to this model were made to follow the evolution of stellar populations and the chemical enrichment by supernovae and asymptotic giant branch (AGB) stars in detail \citep{tornatore2003,tornatore2007a}. It is also critical to model the evolution of supermassive black holes and the resulting AGN feedback to accurately simulate the evolution of the cosmic baryonic content. We follow an updated and expanded upon version \citep{fabjan2010a,hirschmann2014a} of the prescription described by \citet{springel2005d} to evolve black hole sink particles and couple feedback modes to different environmental parameters. All simulations in this work assume a \textit{Planck}-like cosmology\citep[$\Omega_m$=0.307 ; $\Omega_\Lambda$=0.693 ; $H_0$=67.77 km~s$^{-1}$~Mpc$^{-1}$ and $\sigma_8$=0.829, see][]{ade2014}.

\subsection{Cross-identification of structures}
Starting from four catalogs of local supercluster members with galaxy cluster masses \citet{boehringer2021}, \citet{boehringer2021a}, \citet{merluzzi2015} and \citet{monteiro-oliveira2022} B1, B2, M and MO hereafter respectively, we first find the counterparts for the most massive clusters for each region in the catalog by \citet{hernandez-martinez2024a} (the exception here being the Hercules supercluster, where we expand on the catalogue by \citet{hernandez-martinez2024a} by identifying A2147, see \cref{sssec:hercules}). 

In a second step we search for the secondary members using the observed separation of the main and the secondary as a starting point: We first identify all companions for the simulated main cluster, i.e. halos within 50 Mpc/h with masses above $1\times10^{13} (M_\mathrm{vir})$. The search radius is chosen such that the maximal observed companion cluster distance $\mathrm{max}(r_\mathrm{obs})=38.77$ Mpc/h (UGC3355) is contained with a margin within the search region. The mass limit is chosen mainly due to mass resolution limitations. For each observed companion we then search for a match among the simulated companion candidates using the observed relative position of the main cluster and the observed companion. This relative position vector, added to the position of the simulated main halo gives an "ideal" counterpart position with respect to the configuration of the region. The two main criteria for picking a companion match are the deviation from this ideal configuration given by the difference in distance from the main halo and the angular distance to the ideal position and the mass agreement of the companion. Using these criteria aims to ensure that the companion matches we identify are found in a spatial configuration that resembles the observed supercluster structure (in 3D) and mimic the mass distribution in the supercluster. \cref{Virgohal} gives a visual illustration of this process. In the rare cases where the aforementioned criteria yield two or more good candidates we select the objects best matching additional observed properties specifically $M_{500}$ and $L_{X,500}$.  Due to the uncertainties in the observed 3D positions, which are not taken into account in the cross-identification process this can be considered a conservative estimate of how well the regions are reproduced. 

To give an idea of the morphology of the resulting structures we then create mock all-sky observations in X-ray luminosity (0.1-2.4 keV band) and Compton-$y$ using the program SMAC \citep{dolag2005a}. The observer position is chosen according to \citet{dolag2023a}, so all regions are viewed in the same projection. This perspective does not necessarily reflect the selection criterion, which employs the 3D arrangement of the clusters. As an observational reference for the produced Compton-$y$ morphologies we use an Compton-$y$ all-sky map extracted from \textit{Planck} data \citep{aghanim2016} using the MILCA algorithm \citep{hurier2013} and cleaned for regions with high CO emission according to \citet{khatri2016} to which the \textit{Planck} bands used for Compton-$y$ measurements are also sensitive. As an X-ray reference observation, we use the all-sky ROSAT data at 1.5 keV (R6+R7=0.76-2.04 keV) from the ROSAT All-Sky Survey (RASS) \citep{snowden1997}. The map is smoothed with a Gaussian kernel ($\sigma =10$ arcmin). For the regions with a smaller field of view (Shapley, Hercules, Coma) we employ a different smoothing kernel size ($\sigma=5.4$ arcmin) and use R7 (1.05-2.04 keV) only.

\subsection{Quality assessment of the cross-matches}
\label{sec:significance}
  To verify that the cross-identified regions stem from the constraints on the large-scale matter distribution obtained from the local universe rather than the intrinsic statistical clustering properties of the simulation, we introduce an angular two-point significance criterion, expanding on the approach described by \citet{hernandez-martinez2024a}. The basic idea is to obtain an "uncertainty volume" based on the deviation of the position of the simulated cluster from the observed position. Based on this volume, one can then estimate the probability of finding the observed structure randomly by sampling this volume many times in an unconstrained simulation. This is equivalent to performing a null hypothesis test with the supercluster regions obtained from an unconstrained simulation serving as the null hypothesis. To obtain good statistics for the relevant mass range, we use Box0 of the Magneticum suite of simulations. This simulation features a volume of ${2688 \mathrm{\ Mpc/h}^3}$ with a $1.73\times10^{10}M_\odot$ dark matter mass resolution and 5575 haloes with masses exceeding $10^{15}M_\odot$.
  
    \cref{sketch} visualizes how the probabilities are calculated for the pair case: First, the expected position of the less massive companion is calculated by obtaining $r_{obs}$, the relative distance vector of the observed counterpart, and adding this vector to the \textit{simulated} position of the main halo. This expected position ($x_{2,exp}$) together with the position of the simulated companion ($M_{2,sim}$) and the main halo ($M_1$) spans a triangle with the opening angle $\Theta$. The volume corresponding to the spherical volume used for the one-point case by \citet{hernandez-martinez2024a} is then the cone obtained by rotating this triangle around the bisector of $\Theta$. This cone covers a solid angle of $d\Omega \approx \pi\Theta^2$ (for small $\Theta$), which quantifies how close the simulated companion is to the expected position in the reference coordinate system of the main halo. We estimate the probability of finding the companion within $d\Omega$ by sampling random cones centered on randomly selected massive clusters (we choose a threshold of $M_\mathrm{vir}>10^{15}M_\odot$ to sample only the most massive regions which can be expected to have the highest occurrence of massive companions) in the reference simulation. We then count the frequency of companions with a mass equal to or larger than the mass of the SLOW counterpart. In total we sample $10^6$ cones in 5575 main halo vicinities. We further require that the companions lie in the radial range $r_{\mathrm{sim}}-dr_\mathrm{sim,obs}/2<r<r_{\mathrm{sim}}+dr_\mathrm{sim,obs}/2$ to sample the supercluster regions in a similar radial regime. The geometric significance is therefore given by
    \begin{equation}
    \textit{p}_\mathrm{pair}=1-\frac{N_\mathrm{found}}{N_\mathrm{cones}}
    \end{equation}
    the inverse of the fraction of cones sampled that contain at least one companion of $M>M_{2,\mathrm{sim}}$. Because we leave the mass for the randomly sampled companions unbounded towards more massive objects and because we take the observed relative positions of the secondary clusters as ground truth, without accounting for the uncertainties these positions themselves carry, the methods introduced above provide a conservative estimate of how well the simulations reproduce these regions.\footnote{This process can also be understood as defining a spatial metric $\mathcal{D}(r,dr,d\Omega)$. It is then possible to calculate for each cluster $\mathcal{D}_\mathrm{sim}=\mathcal{D}(r_\mathrm{obs},dr_\mathrm{sim,obs},d\Omega)$ and compute $\textit{p}_\mathrm{pair}(\mathcal{D}_\mathrm{sim},M_\mathrm{sim})$ as the fraction of companion clusters of a mass of at least $M_\mathrm{sim}$, where $\mathcal{D}>\mathcal{D}_\mathrm{sim}$ with regards to a randomly sampled observed position in the reference simulation.}
     \begin{figure}[ht]
   \centering
   \includegraphics[width=\columnwidth]{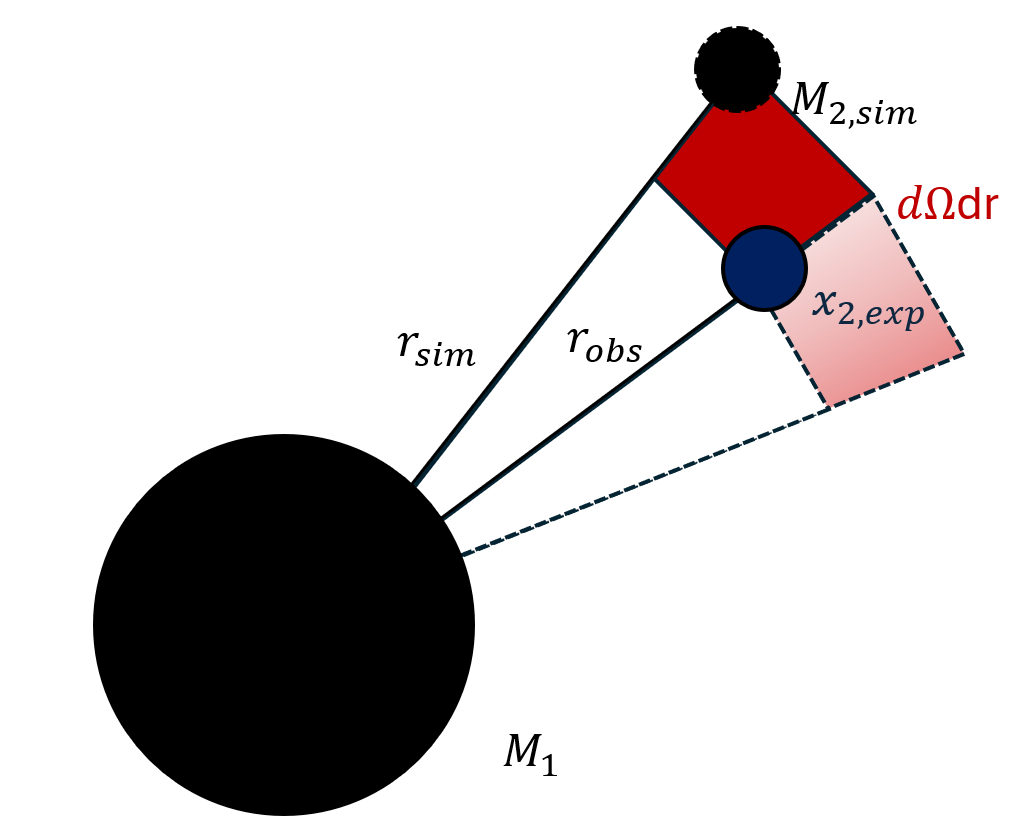}
   \caption{Schematic of how the two-point significances are estimated based on the radial ($dr$) and angular ($d\Omega$) separation of the simulated secondary cluster $M_2$ to the expected position ($x_{2,\mathrm{exp}}$). The red color indicates the part of the cone that is used as a "significance volume" to compute the probability of obtaining the geometric arrangement randomly. }
              \label{sketch}
    \end{figure}

\subsection{Simulating ahead: The Clairvoyant N-body forward simulation suite}
\label{ssec: Clairvoyant}
To analyze to what degree the cross-identified supercluster regions are actually collapsing numerically, we perform forward simulations using a lower resolution ($768^3$) of the SLOW initial conditions set and let the dark matter particles (no gas particles were considered) evolve until a scale factor of $a=1000$, equivalent to 119 Gyrs into the future or equivalent cosmological "redshift" z=-0.999 (it should be noted here that negative redshifts are not physically meaningful but serve as an illustrative temporal parameter), corresponding to roughly $8t_H$ with $t_H=1/H_0$ and our assumed $h=0.6777$. This specific time is chosen to ensure that the rapid expansion of the background due to dark energy domination has frozen in all large-scale evolution. As was already remarked by \citet{nagamine2003a}, this causes the simulation itself to speed up at late times. We then run the \texttt{Subfind} structure finder in an identical setup to the parent simulation on this evolved "final" state in order to identify final haloes.

\section{Results}
\subsection{Superclusters in the SLOW simulations}
\label{ssec:SLOWsups}
As the most massive and galaxy rich assemblages of matter, superclusters can be expected to be among the most well constrained regions in the SLOW simulations: Since the constraints are derived from the galaxy peculiar velocity field, their high galaxy density results in a well-sampled velocity field in these regions. We selected supercluster regions and member clusters from three observational supercluster catalogs and identified the corresponding regions in the SLOW simulations by utilizing the cross-matched most massive clusters in the simulation as described by \citet{hernandez-martinez2024a}. We then identified the companion clusters and groups as outlined in the previous section.
General properties of the 6 supercluster regions for which we find viable counterparts with the methods introduced above are shown in \cref{table:set}. As can be seen from the table we are able to identify counterparts for most of the massive supercluster members in the observational catalogs. \cref{Mollweide} shows the distribution of these regions on a simulated full-sky Compton-$y$ map generated with SMAC \citep{dolag2005a} and centered on the SLOW optimal center from \citet{dolag2023a}. The colored dots indicate each supercluster respectively and show the simulated superclusters to reproduce the general spatial arrangement of the observed regions.

The cross matched member clusters are listed in \cref{membertable} arranged by region and with their most important properties. To provide more details on each region and the exact arrangement of the member clusters, we show a four-panel figure for each supercluster - \cref{Shapley} for the Shapley supercluster and \cref{P-P} for Perseus-Pisces. The remaining four regions can be found in \cref{AppendixB}. The four panel plots are structured as follows: The top row (panels A and B) compares a mock image from the simulation (A) and the respective X-ray observation (B). To give a better idea of the underlying structure, panel A shows the gas density contours of the mock region in white (2D gas overdensity levels of $[1.25,10]*\bar{\rho}_{\mathrm{image}}$). The density was calculated using SMAC using a suitable projection depth to capture all of the member clusters. In the bottom row we show the predicted Lagrangian \textbf{collapse volume} of the supercluster calculated from the forward simulation described in \cref{ssec: Clairvoyant} at different points in time. This Lagrangian volume is given by the Friends-of-Friends (FOF) group at the last snapshot (at z=-0.999). In the left panel (C) it is shown in the initial conditions with the individual member Lagrangian volumes overplotted. In the right panel (D) the Lagrangian volume is shown at z=0 in the context of the surrounding large scale structure.
Note that it is primarily the linear scale of the density/velocity field that is constrained by the observational data for SLOW. Additionally resolution effects from the simulation itself can already impact haloes in this mass regime even if they are induced in the correct place by the constraints.  Therefore we do not expect the low mass member clusters and groups ($M \approx 10^{13}M_\odot$) to be reproduced accurately and focus on the most massive member clusters for the cross identification. 
\begin{table*}[h!]

\centerline{
\begin{tabular}{llllll}
Region         & Reference & \makecell[l]{Observed members $M_\mathrm{vir}\geq10^{14}M_\odot$\\ (all members)} & Counterparts& \makecell[l]{Maximum Influence radius\\ $\mathrm{[Mpc/h]}$ at z=0} \\
\hline
Shapley        & M         &   4(11)&    6&15.95\\
Perseus-Pisces & B2         &  7(22)&  8&9.90\\
Coma           & B1         & 5(13)        &   4&13.93\\
Virgo          & B1         &  1(5)& 3&8.70\\
Centaurus      & B1         &  2(10)&   4&12.71\\
Hercules(core) & MO&2(4)&3&10.94
\end{tabular}
}
\caption{Reference set and cross identified member clusters. Additionally we give the maximum extent of the collapse region at z=0.}
\label{table:set}
\end{table*}
\begin{figure*}[ht]
   \centering
   \includegraphics[width=0.49\textwidth]{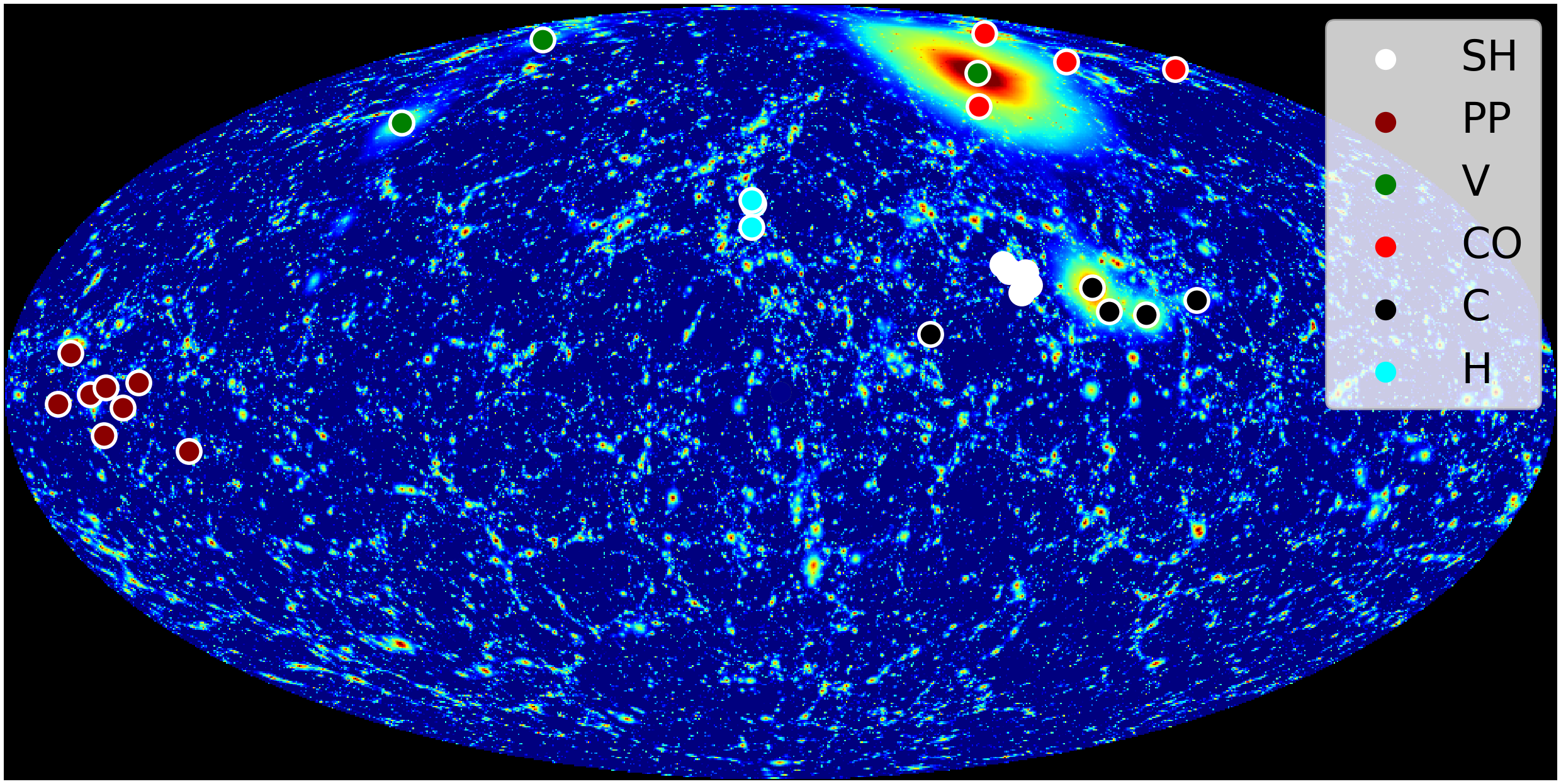}
   \includegraphics[width=0.49\textwidth]{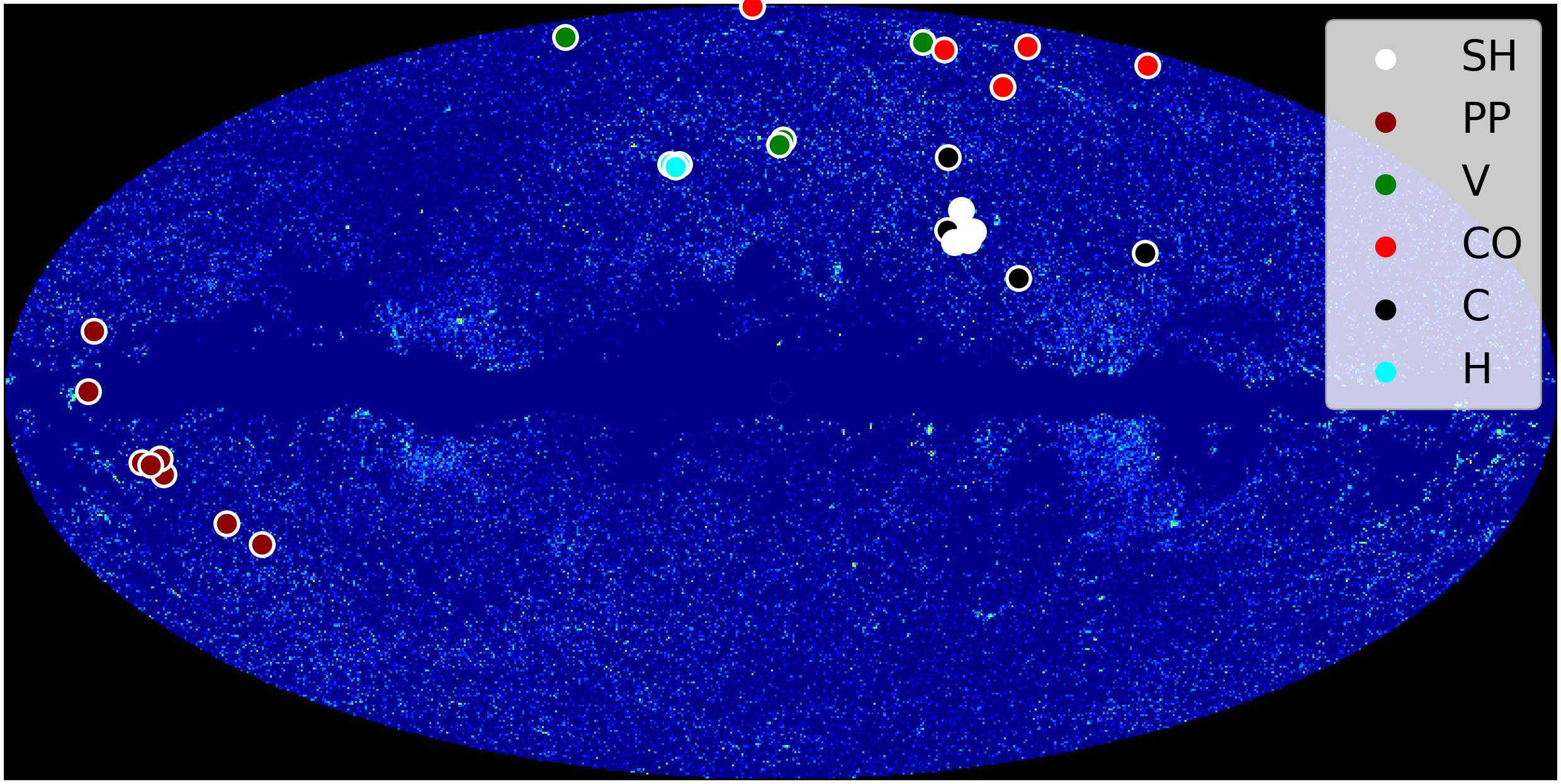}
   \caption{Simulated mock Compton-$y$ map from SLOW (left) and observed Compton-$y$ map (right) from Planck \citep{ade2014}. SLOW cross identified supercluster members compared to selected observationally obtained supercluster regions using the catalogues from \citet{boehringer2021, boehringer2021a,proust2006,monteiro-oliveira2022}. The different colors indicate the regions: C=Centaurus, CO=Coma, H=Hercules, PP=Perseus-Pisces, SH=Shapley, V=Virgo (Local supercluster)}
              \label{Mollweide}
\end{figure*}

\begin{table*}[h!]
\tiny
\centerline{
\begin{tabular*}{1.0223\textwidth}{|c|lll|lll|ll|ll|ll|ll|c|}
\hline
\multirow{2}{*}{\makecell[l]{Cluster/\\Group}} & \multicolumn{3}{l|}{$\mathrm{SG}_{\mathrm {sim}}$} & \multicolumn{3}{l|}{$\mathrm{SG}_{\mathrm {obs}}$} & \multicolumn{2}{l|}{$M_\mathrm{200}$} & \multicolumn{2}{l|}{$L_{X}$} & \multicolumn{2}{l|}{$M_\mathrm{vir}$} & \multicolumn{2}{l|}{$M_{500}$}& \multirow{2}{*}{\textit{p}} \\ \cline{2-15} 
 & \multicolumn{1}{l|}{X} & \multicolumn{1}{l|}{Y} & \multicolumn{1}{l|}{Z}  & \multicolumn{1}{l|}{X} & \multicolumn{1}{l|}{Y} & \multicolumn{1}{l|}{Z} & \multicolumn{1}{l|}{Sim} & \multicolumn{1}{l|}{Obs} & \multicolumn{1}{l|}{Sim} & \multicolumn{1}{l|}{Obs} & \multicolumn{1}{l|}{Sim} & Obs & \multicolumn{1}{l|}{Sim} & \multicolumn{1}{l|}{Obs}& \\ \hline \hline
\cellcolor{blue!6}\Centerstack[l]{Shapley\\A3558} & \multicolumn{1}{l|}{-141.91} & \multicolumn{1}{l|}{66.10} & -37.17 & \multicolumn{1}{l|}{-130.29} & \multicolumn{1}{l|}{78.70} & -3.64 & \multicolumn{1}{l|}{13.07} & - & \multicolumn{1}{l|}{13.25} & $17.4^\dag$ & \multicolumn{1}{l|}{17.38} & 37.10 & \multicolumn{1}{l|}{9.32} & \makecell[l]{4.41\\$\pm$0.273}& - \\ \hline
\cellcolor{blue!6} A3571 & \multicolumn{1}{l|}{-142.64} & \multicolumn{1}{l|}{54.10} & -33.88 & \multicolumn{1}{l|}{-109.39} & \multicolumn{1}{l|}{58.57} & 4.39 & \multicolumn{1}{l|}{5.93} & - & \multicolumn{1}{l|}{5.45} & \makecell[l]{${21.2}^\dag$\\$\pm$0.3} & \multicolumn{1}{l|}{7.09} & 8.76 & \multicolumn{1}{l|}{4.04} & \makecell[l]{4.67\\$\pm$ 0.21}&0.030 \\ \hline
\cellcolor{blue!6} A3560 & \multicolumn{1}{l|}{-145.74} & \multicolumn{1}{l|}{60.55} & -39.70 & \multicolumn{1}{l|}{-130.96} & \multicolumn{1}{l|}{73.42} & -2.58 & \multicolumn{1}{l|}{5.55} & - & \multicolumn{1}{l|}{4.23} & \makecell[l]{$4.17^\dag$\\$\pm0.17$\\} & \multicolumn{1}{l|}{6.88} & $3.40^\flat$ & \multicolumn{1}{l|}{3.21} & -& 0.002\\ \hline
A1736 & \multicolumn{1}{l|}{-137.66} & \multicolumn{1}{l|}{65.92} & -22.87 & \multicolumn{1}{l|}{-117.47} & \multicolumn{1}{l|}{83.34} & -0.73 & \multicolumn{1}{l|}{4.74} & - & \multicolumn{1}{l|}{3.65} & \makecell[l]{$8.51^\dag$\\ $\pm$0.87} & \multicolumn{1}{l|}{6.68} & 18.20 & \multicolumn{1}{l|}{3.28} & \makecell[l]{1.63\\+0.46\\-0.52}&0.003 \\ \hline
A3559 & \multicolumn{1}{l|}{-134.88} & \multicolumn{1}{l|}{61.53} & -24.59 & \multicolumn{1}{l|}{-121.23} & \multicolumn{1}{l|}{78.36} & -0.93 & \multicolumn{1}{l|}{2.30} & - & \multicolumn{1}{l|}{0.79} & - & \multicolumn{1}{l|}{3.14} & $0.2^\flat$ & \multicolumn{1}{l|}{1.57} & -&0.04 \\ \hline
\Centerstack[l]{A3532\\(+A3530)} & \multicolumn{1}{l|}{ -158.49} & \multicolumn{1}{l|}{89.78} &  -55.89 & \multicolumn{1}{l|}{-141.76} & \multicolumn{1}{l|}{ 101.20} & -27.98 & \multicolumn{1}{l|}{4.22} & - & \multicolumn{1}{l|}{2.04} & \Centerstack[l]{${0.08}^\dag$ \\ $\pm  0.17$} & \multicolumn{1}{l|}{5.28} & \Centerstack[l]{8.9\\ $\pm$ 4.4 } & \multicolumn{1}{l|}{2.68} & -&0.000 \\ \hline\hline
\cellcolor{blue!6}\Centerstack[l]{Perseus\\A426} & \multicolumn{1}{l|}{59.22} & \multicolumn{1}{l|}{2.13} & -24.58 & \multicolumn{1}{l|}{49.94} & \multicolumn{1}{l|}{-10.73} & -12.98 & \multicolumn{1}{l|}{8.07} & 13.66 & \multicolumn{1}{l|}{5.81} & 37.88 & \multicolumn{1}{l|}{10.63} & 16.3 & \multicolumn{1}{l|}{5.10} & -&- \\ \hline
\cellcolor{blue!6} AWM 7 & \multicolumn{1}{l|}{60.08} & \multicolumn{1}{l|}{-6.71} & -23.32 & \multicolumn{1}{l|}{49.90} & \multicolumn{1}{l|}{-13.20} & -9.41 & \multicolumn{1}{l|}{0.73} & 3.72 & \multicolumn{1}{l|}{0.08} & 4.64 & \multicolumn{1}{l|}{0.90} & 4.8 & \multicolumn{1}{l|}{0.53} & -& 0.127 \\ \hline
\Centerstack[l]{UGC\\2562} & \multicolumn{1}{l|}{101.39} & \multicolumn{1}{l|}{-2.36} & -27.86 & \multicolumn{1}{l|}{85.97} & \multicolumn{1}{l|}{-19.41} & -19.40 & \multicolumn{1}{l|}{0.80} & 0.45 & \multicolumn{1}{l|}{0.81} & 0.15 & \multicolumn{1}{l|}{1.15} & 1.81 & \multicolumn{1}{l|}{0.42} & -&0.019 \\ \hline
\makecell[l]{CIZAJ\\0300.7\\+4427} & \multicolumn{1}{l|}{ 81.61} & \multicolumn{1}{l|}{ -21.57} &  -39.01 & \multicolumn{1}{l|}{85.13} & \multicolumn{1}{l|}{-17.63} & -14.87 & \multicolumn{1}{l|}{1.66} & 3.70 & \multicolumn{1}{l|}{0.86} & 4.65 & \multicolumn{1}{l|}{2.09} & 8.4 & \multicolumn{1}{l|}{1.15} & -& 0.008 \\ \hline
3C129 & \multicolumn{1}{l|}{55.55} & \multicolumn{1}{l|}{11.33} & -28.56 & \multicolumn{1}{l|}{56.90} & \multicolumn{1}{l|}{2.82} & -24.05 & \multicolumn{1}{l|}{1.82} & 3.41 & \multicolumn{1}{l|}{0.80} & 4.07 & \multicolumn{1}{l|}{2.29} & $7.80^\ddag$ & \multicolumn{1}{l|}{0.86} & - &0.097\\ \hline
A262 & \multicolumn{1}{l|}{64.51} & \multicolumn{1}{l|}{-12.44} & -2.90 & \multicolumn{1}{l|}{42.41} & \multicolumn{1}{l|}{-19.81} & -1.60 & \multicolumn{1}{l|}{1.21} & 2.26 & \multicolumn{1}{l|}{0.75} & 2.06 & \multicolumn{1}{l|}{1.38} & 2.82 & \multicolumn{1}{l|}{0.93} & -&0.215 \\ \hline
NGC507 & \multicolumn{1}{l|}{49.14} & \multicolumn{1}{l|}{-22.50} & -32.42 & \multicolumn{1}{l|}{40.34} & \multicolumn{1}{l|}{-23.03} & 2.24 & \multicolumn{1}{l|}{1.72} & 1.20 & \multicolumn{1}{l|}{0.35} & 0.73 & \multicolumn{1}{l|}{2.04} & $1.10^\ddag$ & \multicolumn{1}{l|}{1.27} & - & 0.194\\ \hline
\Centerstack[l]{UGC\\3355} & \multicolumn{1}{l|}{95.07} & \multicolumn{1}{l|}{1.84} & -55.58 & \multicolumn{1}{l|}{69.62} & \multicolumn{1}{l|}{18.69} & -28.81 & \multicolumn{1}{l|}{1.58} & 1.18 & \multicolumn{1}{l|}{0.47} &0.73& \multicolumn{1}{l|}{2.04} & 3.32 & \multicolumn{1}{l|}{1.10} & -&0.32 \\ \hline \hline
 \cellcolor{blue!6} \Centerstack[l]{Coma\\A1656} & \multicolumn{1}{l|}{-2.93} & \multicolumn{1}{l|}{82.27} & -9.14 & \multicolumn{1}{l|}{0.48} & \multicolumn{1}{l|}{72.79} & 10.59 & \multicolumn{1}{l|}{15.17} & 8.53 & \multicolumn{1}{l|}{14.71} & 17.79 & \multicolumn{1}{l|}{18.60} & 15.9 & \multicolumn{1}{l|}{9.50} & \makecell[l]{5.29\\$\pm$ 0.20}&- \\ \hline
\cellcolor{blue!6}A1367 & \multicolumn{1}{l|}{-2.89} & \multicolumn{1}{l|}{71.35} & -18.73 & \multicolumn{1}{l|}{-2.94} & \multicolumn{1}{l|}{68.62} & -12.68 & \multicolumn{1}{l|}{1.67} & 2.55 & \multicolumn{1}{l|}{0.58} & 2.53 & \multicolumn{1}{l|}{2.18} & 5.4 & \multicolumn{1}{l|}{1.08} & \makecell[l]{1.76\\$\pm$ 0.14}& 0.106\\ \hline
A1185 & \multicolumn{1}{l|}{24.59} & \multicolumn{1}{l|}{112.10} & -27.24 & \multicolumn{1}{l|}{16.31} & \multicolumn{1}{l|}{99.93} & -25.16 & \multicolumn{1}{l|}{3.35} & 1.22 & \multicolumn{1}{l|}{2.49} & 0.78 & \multicolumn{1}{l|}{4.37} & 11.8 & \multicolumn{1}{l|}{2.01} & \makecell[l]{1.27\\$\pm$ 0.19}& 0.012\\ \hline
MKW4 & \multicolumn{1}{l|}{-27.22} & \multicolumn{1}{l|}{51.95} & -16.53 & \multicolumn{1}{l|}{-22.11} & \multicolumn{1}{l|}{57.35} & -12.78 & \multicolumn{1}{l|}{0.80} & 1.32 & \multicolumn{1}{l|}{0.86} & 0.88 & \multicolumn{1}{l|}{0.90} & 0.618 & \multicolumn{1}{l|}{0.57} & -& 0.19 \\ \hline \hline
\cellcolor{blue!6}Virgo & \multicolumn{1}{l|}{-3.60} & \multicolumn{1}{l|}{10.36} & -1.64 & \multicolumn{1}{l|}{-3.37} & \multicolumn{1}{l|}{14.71} & -0.66 & \multicolumn{1}{l|}{8.3} & 1.64 & \multicolumn{1}{l|}{7.13} & 0.88 & \multicolumn{1}{l|}{9.8} & \makecell[l]{6.3 \\$\pm$ 0.9} & \multicolumn{1}{l|}{6.49} & \makecell[l]{4.76\\ $\pm$ 0.55}&- \\ \hline
\cellcolor{blue!6} \cellcolor{blue!6}\Centerstack[l]{M94\\group} & \multicolumn{1}{l|}{5.46} & \multicolumn{1}{l|}{7.51} & 2.34 & \multicolumn{1}{l|}{1.24} & \multicolumn{1}{l|}{5.08} & 0.87 & \multicolumn{1}{l|}{0.45} & - & \multicolumn{1}{l|}{0.14} &  & \multicolumn{1}{l|}{0.52} & 0.03& \multicolumn{1}{l|}{0.36} & - & 0.036 \\ \hline
\Centerstack[l]{NGC\\5846} & \multicolumn{1}{l|}{-20.25} & \multicolumn{1}{l|}{1.19} & 22.86 & \multicolumn{1}{l|}{-9.16} & \multicolumn{1}{l|}{13.29} & 10.08 & \multicolumn{1}{l|}{0.13} & 0.36 & \multicolumn{1}{l|}{0.08} & 0.05 & \multicolumn{1}{l|}{0.46} & 0.84 & \multicolumn{1}{l|}{0.23} & $0.14^{*}$ &0.152 \\ \hline
\hline
\cellcolor{blue!6} \Centerstack[l]{Centaurus\\/A3526} & \multicolumn{1}{l|}{-22.82} & \multicolumn{1}{l|}{11.21} & -13.26 & \multicolumn{1}{l|}{-34.25} & \multicolumn{1}{l|}{14.93} & -7.56 & \multicolumn{1}{l|}{8.58} & 3.27 & \multicolumn{1}{l|}{10.05} & \makecell[l]{3.76} & \multicolumn{1}{l|}{10.07} & 10.8 & \multicolumn{1}{l|}{6.17} & \makecell[l]{1.23\\$\pm$ 0.11}&- \\ \hline
\cellcolor{blue!6}\Centerstack[l]{NGC\\4936} & \multicolumn{1}{l|}{-15.97} & \multicolumn{1}{l|}{11.66} & -23.04 & \multicolumn{1}{l|}{-28.37} & \multicolumn{1}{l|}{18.76} & -3.56 & \multicolumn{1}{l|}{0.91} & 0.44 & \multicolumn{1}{l|}{0.33} & 0.15 & \multicolumn{1}{l|}{1.18} & $0.23^\S$ & \multicolumn{1}{l|}{0.63} & -&0.288 \\ \hline
AS0753 & \multicolumn{1}{l|}{-42.15} & \multicolumn{1}{l|}{7.50} & 6.38 & \multicolumn{1}{l|}{-37.60} & \multicolumn{1}{l|}{18.51} & 3.53 & \multicolumn{1}{l|}{0.82} & 0.52 & \multicolumn{1}{l|}{0.65} &0.20& \multicolumn{1}{l|}{1.01} & 1.65 & \multicolumn{1}{l|}{0.56} & - & 0.356\\ \hline
A3574E & \multicolumn{1}{l|}{-35.04} & \multicolumn{1}{l|}{14.30} & -23.29 & \multicolumn{1}{l|}{-43.45} & \multicolumn{1}{l|}{25.86} & 3.00 & \multicolumn{1}{l|}{2.27} & 0.60 & \multicolumn{1}{l|}{1.12} &0.25& \multicolumn{1}{l|}{2.78} & 1.93 & \multicolumn{1}{l|}{1.64} & $0.32^{*}$&0.049 \\ \hline \hline
 \cellcolor{blue!6} \Centerstack[l]{Hercules\\A2147} & \multicolumn{1}{l|}{-77.42} & \multicolumn{1}{l|}{77.33} & 100.57 & \multicolumn{1}{l|}{-26.06} & \multicolumn{1}{l|}{71.85} & 86.49 & \multicolumn{1}{l|}{13.02} & \makecell{13.5\\+2.1\\-1.7} & \multicolumn{1}{l|}{14.44} &\makecell[l]{${7.49}^\dag$\\± 0.40}  & \multicolumn{1}{l|}{15.3} & 53.1 & \multicolumn{1}{l|}{9.82} & $2.51^{*}$&\textcolor{violet}{0.049}\\ \hline
 \cellcolor{blue!6} A2151 & \multicolumn{1}{l|}{-77.40} & \multicolumn{1}{l|}{69.12} & 100.58 & \multicolumn{1}{l|}{-22.47} & \multicolumn{1}{l|}{71.88} & 87.34 & \multicolumn{1}{l|}{1.65} & \makecell[l]{2.88\\+0.31\\-0.27} & \multicolumn{1}{l|}{0.80} & \makecell[l]{$2.33^\dag$\\$\pm$0.07} & \multicolumn{1}{l|}{1.90} & 2.88 & \multicolumn{1}{l|}{1.21} & $1.38^{*}$& 0.136 \\ \hline
  \cellcolor{blue!6}A2152 & \multicolumn{1}{l|}{-76.53} & \multicolumn{1}{l|}{79.80} & 101.63 & \multicolumn{1}{l|}{-27.38} & \multicolumn{1}{l|}{78.01} & 96.31 & \multicolumn{1}{l|}{0.61} & \makecell[l]{0.74\\+0.13\\-0.10} & \multicolumn{1}{l|}{0.70} & \makecell[l]{$0.64^{*}$} & \multicolumn{1}{l|}{0.81} & 0.72 & \multicolumn{1}{l|}{0.43} & $0.6^{*}$& 0.202 \\ \hline
\end{tabular*}
}
\caption{Supercluster members and their properties compared to observations. All masses are in units of $10^{14}M_\odot$ and X-ray Luminosities are given in units of $10^{43}erg/s/h^2$, measured in the 0.1 to 2.4 keV band if not specified otherwise. The base catalog for these properties are B,B1,M and MO. Where data was incomplete we used \citep{piffaretti2011} (*) and \citep{ikebe2002} ($\dag$) to supplement information. The dynamical masses are obtained from \citet{tully2015} with supplementary data from \citet{pinzke2011} ($\ddag$), \citet{mendel2008} ($\lozenge$), \citet{makarov2011} ($\S$), M ($\flat$) and MO (Hercules supercluster). Each subsection of the table lists the most massive or main member first. The blue shading in the names column indicates whether the member is bound to the supercluster core according to the method described by \cref{collapse}. The probabilities (\textit{p}) in the last column represent the probability that a given supercluster member is identified purely by chance as introduced in \cref{sec:significance}. The main halos of the regions do not have an associated probability with this method. We refer to \citet{hernandez-martinez2024a} for the significance of these matches. The probability for A2147, the only main halo in this work not identified by \citet{hernandez-martinez2024a} is computed according to the method outlined in that study, see also \cref{sssec:hercules}. This is indicated by the \textcolor{violet}{violet} color.}
\label{membertable}
\end{table*}
\subsubsection{The Shapley supercluster region}
   \begin{figure*}[ht]
   \centering
   \includegraphics[width=\textwidth]{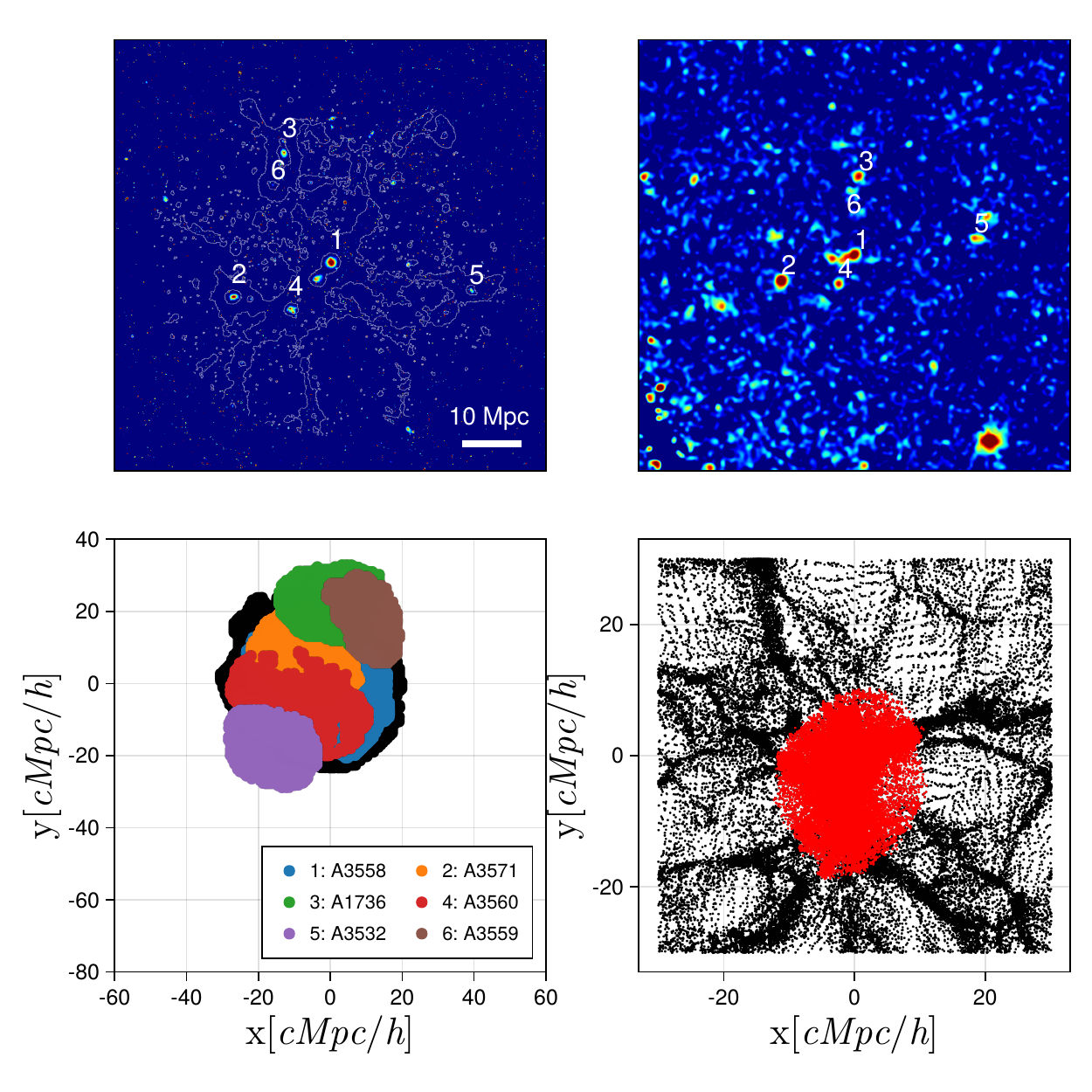}
   \caption{\textit{Top left:} Shapley supercluster region in the SLOW simulation with the cross identified counterparts: 1:A3558, 2:A3571, 3:A1736, 4: A3560, 5:A3532, 6:A3559. The white bar indicates 10 Mpc at the distance of the main cluster. The white contours show the 2D gas overdensity levels of $[1.25,10]*\bar{\rho_{\mathrm{image}}}$. \textit{Top right:} The same FOV as viewed by ROSAT. \textit{Bottom left:} Supercluster core region (black) according to the Clairvoyant N-Body forward simulation in the initial conditions ($z=120$) with the member clusters from \cref{membertable} overplotted as colored dots. \textit{Bottom right:} The collapse region (red) and its environment at $z=0$.}
              \label{Shapley}
    \end{figure*}
   \begin{figure*}[ht]
   \centering
   \includegraphics[width=0.49\textwidth]{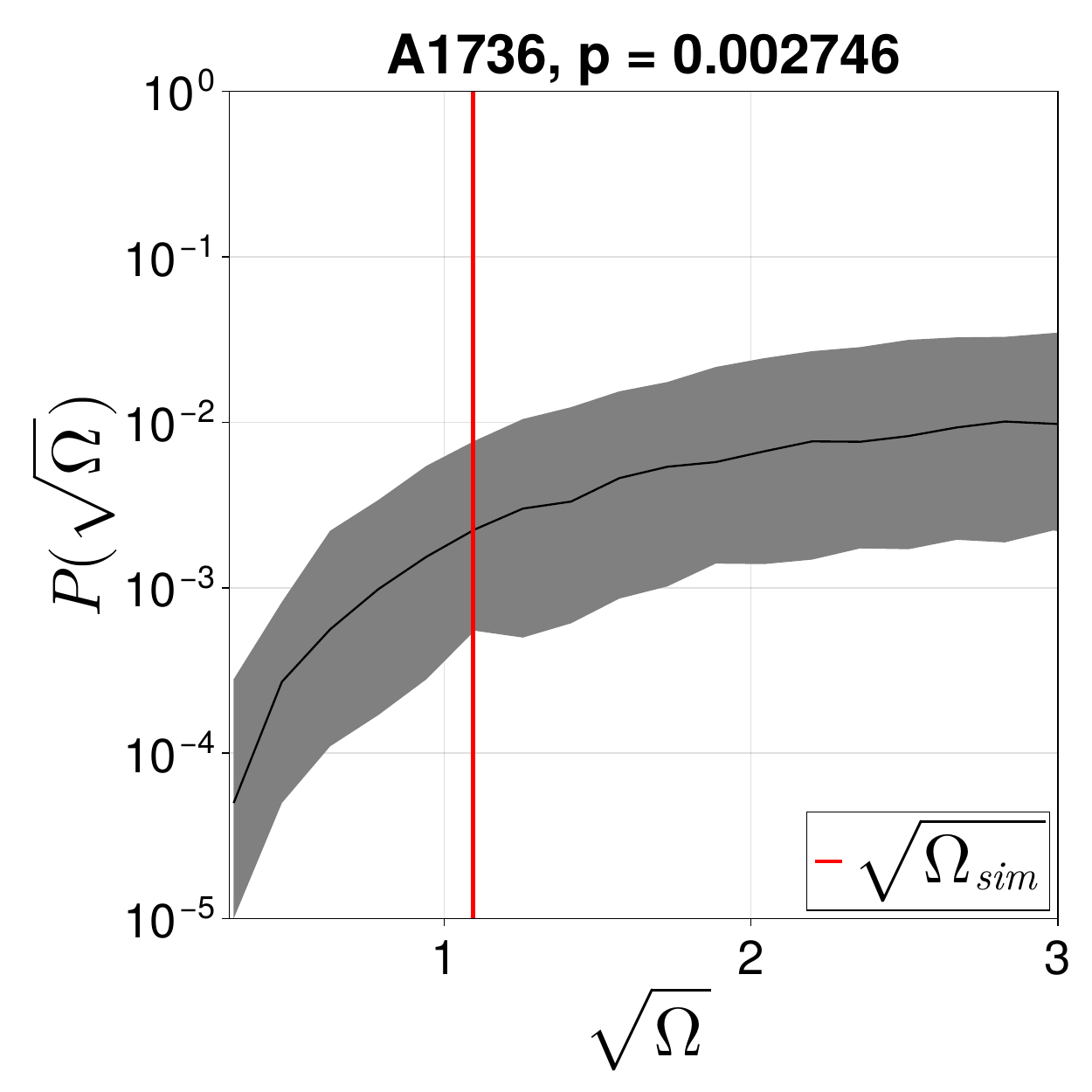}
   \includegraphics[width=0.49\textwidth]{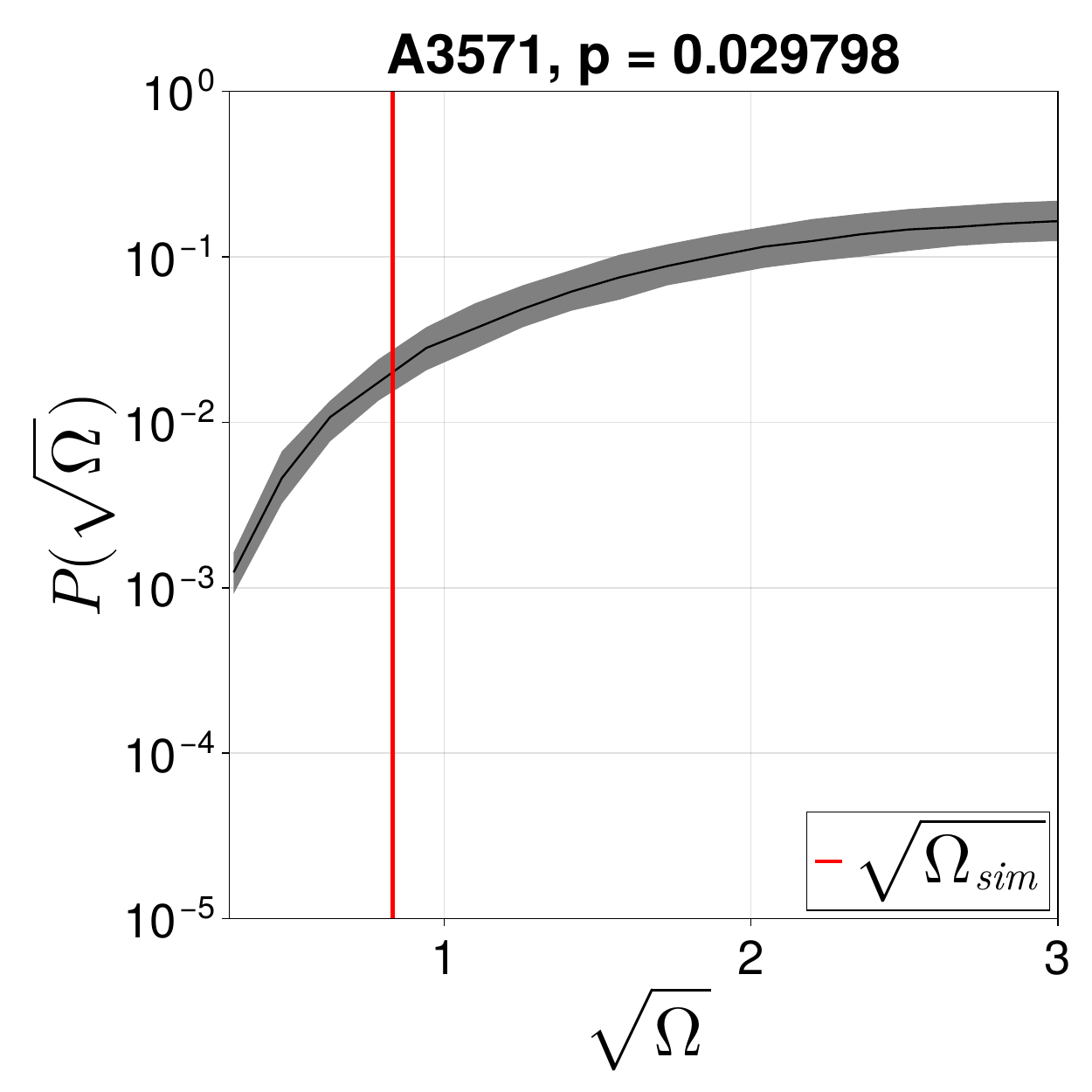}
   \includegraphics[width=0.49\textwidth]{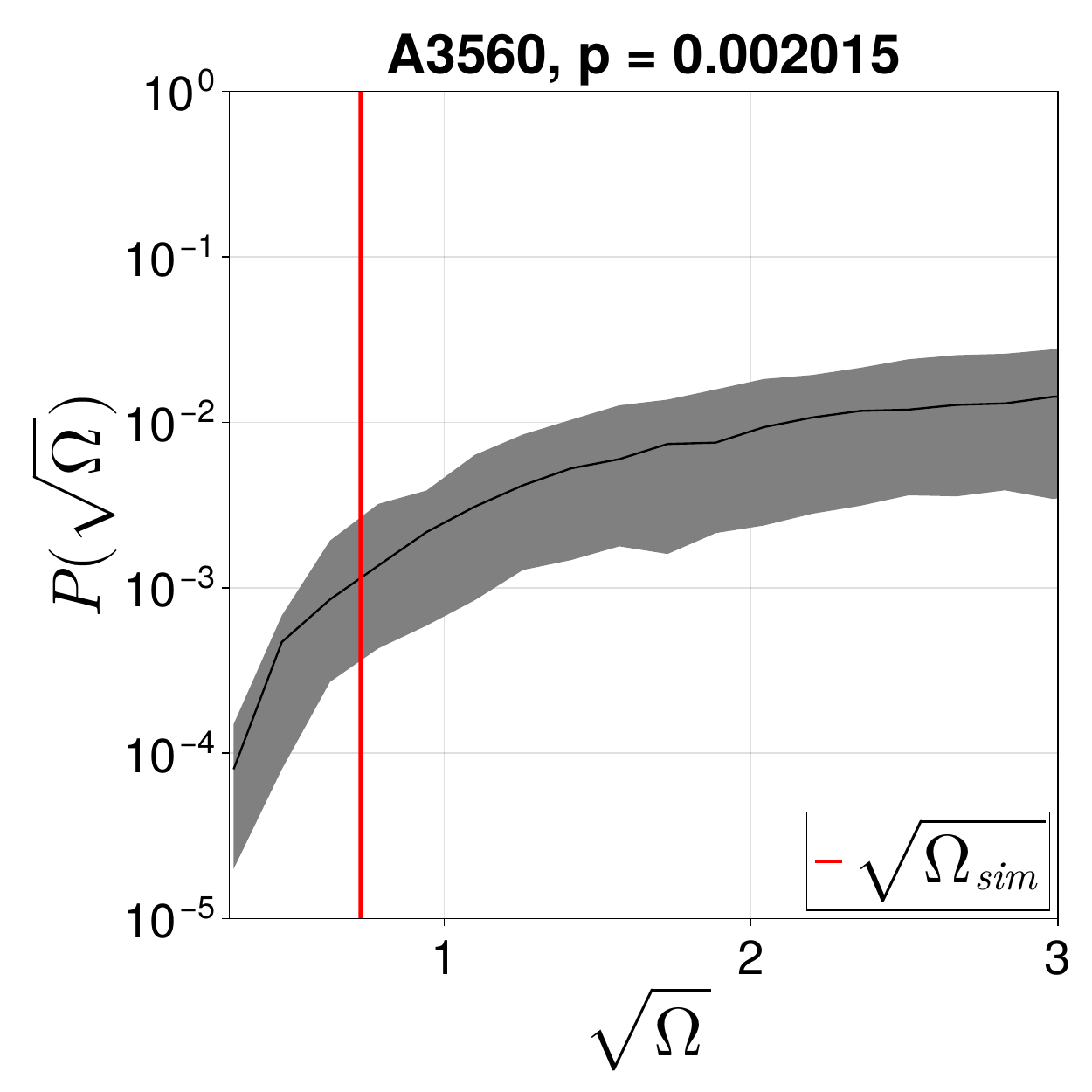}
   \includegraphics[width=0.49\textwidth]{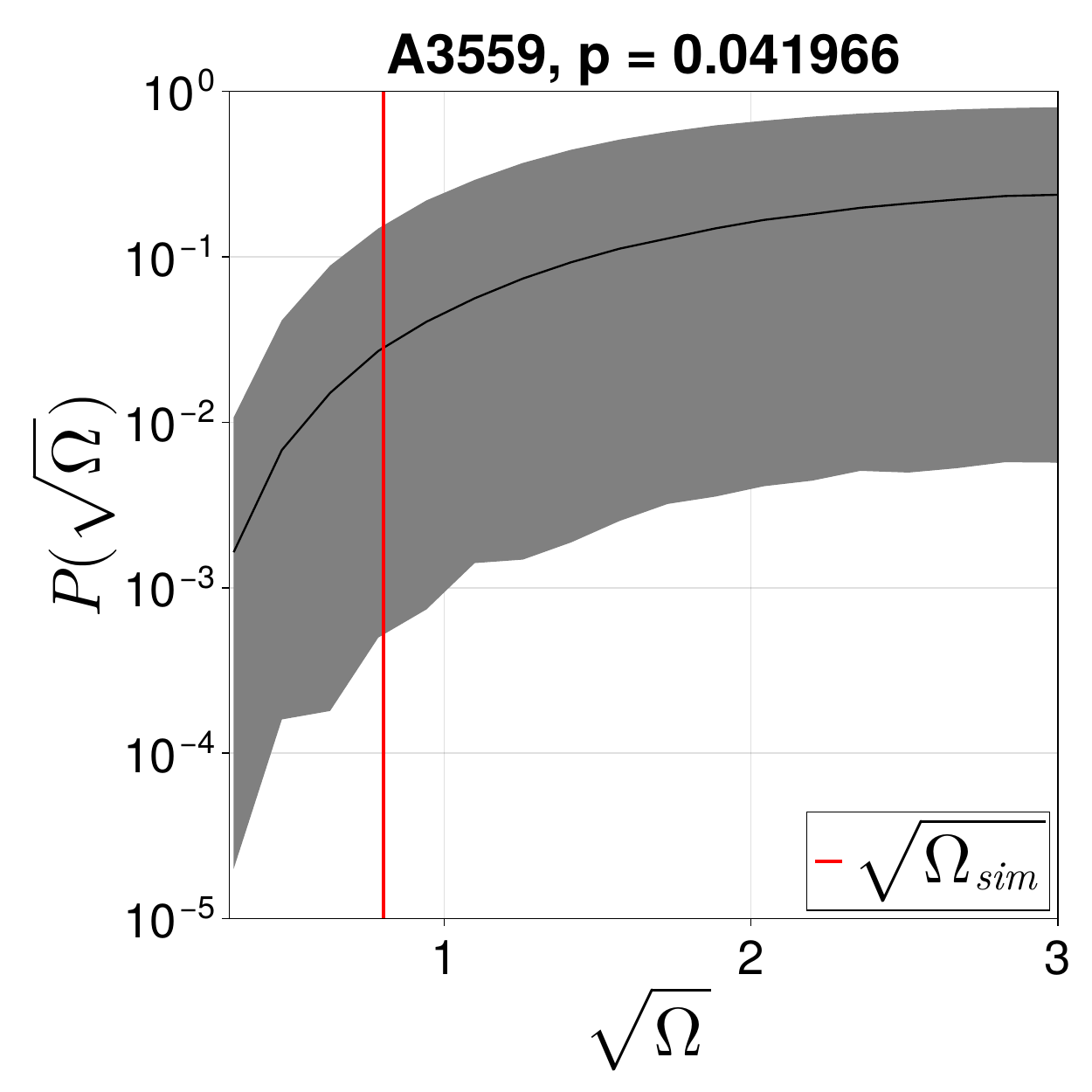}
   \caption{Two-point probabilities for the Shapley member clusters as a function of the opening angle $\sqrt{\Omega}$ of the cone spanned by the observed and simulated relative position of the secondary cluster to the main. The grey shaded regions reflect the mass uncertainty of the secondary halo. The red vertical line indicates the deviation angle measured in the SLOW simulation. The x value of the intersect of this line with the angular probability function gives the probability of finding the secondary within the deviation cone in a random simulation.}
              \label{sig1}
    \end{figure*}
 The Shapley supercluster as identified by \citet{merluzzi2015} consists of 4 massive core members: A3558, A3556, A3562 and A3560. As can be seen in panels A and B of \cref{Shapley}, we recover A3558 and A3560 (numbers 1 and 4 in the plots) in consistent positions in SLOW. Additionally, we find three more members of the outer supercluster, namely A3571, A1736 from \citet{proust2006} and A3559 which is also identified by \citet{merluzzi2015} but less massive. Shapley is one of the most massive supercluster structures observed to date \citep{zucca1993}. This region, especially its central core has been the focus of many multi-wavelength studies, investigating radio and X-ray evidence for sloshing at the center of A3560 \citep{venturi2013}, non-thermal signatures of group-cluster interaction in the region between A3562 and A3558 \citep{venturi2022} as well as optical and spectroscopic analysis of substructures and filaments connecting the supercluster members \citep{merluzzi2015}. A recent study performed a blind extraction of the full filamentary network from the galaxy distribution and connected the star formation activity of the galaxies in the region to the supercluster environment and its filamentary structure \citep{aghanim2024}. The relatively large distance of the Shapley supercluster with a redshift range of $z \in [0.04,0.055]$ \citep{proust2006} puts it at the very edge of the SLOW region one could reasonably expect to be constrained. Evaluating the tracer galaxy positions from Cosmicflows-2, we find only a single galaxy velocity within the collapse volume of the region (see also \cref{Shapleytrace}).  Nevertheless, this structure is recovered to striking detail. The characteristic large-scale arrangement can be clearly seen in \cref{Shapley}, with the gas density contours indicating the central filamentary structure connecting A3558 (1) to A3560 (4) and A3571 (2) as found by \citet{aghanim2024}. Additionally, the structure connecting A3559 (6) and A1736 (3), which is also identified as a filament by \citet{aghanim2024}, can be seen in the underlying gas density. In addition to these morphological agreements, \cref{sig1} shows the two-point significances discussed in \cref{sec:significance} for four exemplary member clusters. The small separation angles ($\sqrt{\Omega_{sim}}$, (red vertical lines) for these members (in combination with their high masses) lead to a very low probability of random detection, quantitatively solidifying the agreement from the projection in \cref{Shapley}. 

A notable distinction from the observed supercluster is the stage of the merger in the core of the region: A3558 and A3562 have already merged at $z=0$ in the simulation forming a double core that is already identified as a single halo and additionally undergoes another merger briefly before z=0 (likely responsible for the lack of a counterpart for A3556). These slight offsets in merging state/time occur quite often and most likely arise from the positional uncertainties that can significantly shift the infall history in collapsing systems. Additional uncertainty factors for these histories are introduced by the assumed background cosmology, the linking criterion of the halo finder, and possibly the exact numerical implementation of the simulation code.
\paragraph{A3532: Differences between one-point and two point results}
The distant member cluster A3532 illustrates how considering the relative positions of the member clusters - as opposed to the distance to their observed position - can impact the cross-identification of SLOW clusters with observed counterparts. The match identified by \citet{hernandez-martinez2024a} is selected because its mass in combination with the small distance to the observed position of A3532 gives it a high one-point significance. Relative to the Shapley SC core A3558, the previous counterpart is, however, in the wrong position, indicating another simulated cluster as a better counterpart based on the arrangement of the wider region. The previous match is found to be a better fit for A3559 with respect to the supercluster structure. \cref{Shapley} shows the situation. Cluster 6 is the previous match, now identified with A3559 and cluster 5 the new counterpart for A3532. The resulting two-point significances for the two matches can also be seen in \cref{sig2}, quantitatively demonstrating the alternative match to be in a better position w.r.t. the supercluster environment. This is the only supercluster member also identified by \citet{hernandez-martinez2024a} where the supercluster environment indicates a different counterpart to the method used in that work, so a disagreement between the two methods appears to be a rare occurrence and can likely be attributed to the previously discussed sparsity of constraints in this particular region.

\subsubsection{The Perseus-Pisces supercluster region}
\begin{figure*}[ht]
   \centering
   \includegraphics[width=\textwidth]{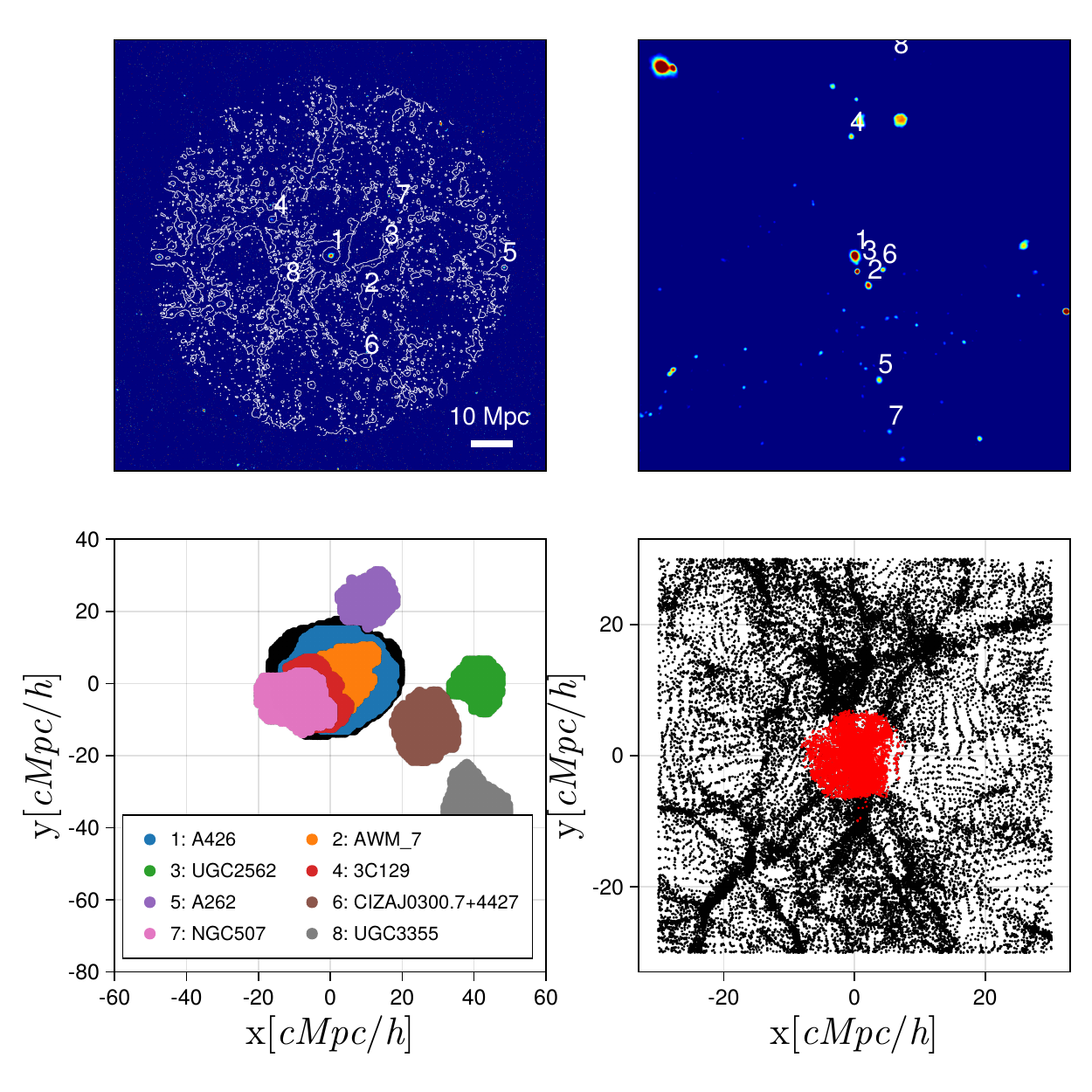}
   \caption{The Perseus-Pisces supercluster region in the SLOW simulation with the cross-identified counterparts: 1: Perseus (A426), 2: AWM 7, 3: UGC2562, 4: 3C129, 5: A262, 6: CIZAJ0300.7+4427, 7: NGC507, 8: UGC3355. The white contours show the 2D gas overdensity levels of $[1.25,10]*\bar{\rho_{\mathrm{image}}}$. \textit{Top right:} The same region as viewed by ROSAT. \textit{Bottom left:} Supercluster core region according to the Clairvoyant N-Body forward simulation in the initial conditions ($z=120$).\textit{Bottom right:} The collapse region (red) and its environment at $z=0$.}
              \label{P-P}
    \end{figure*}
The Perseus-Pisces complex, as identified by \citet{boehringer2021}, is notable by its filamentary and branching morphology \citep{einasto1980} and is comprised of 7 high mass members, namely A426, AWM7, CIZAJ0300.7+4427, 3C129, A262, NGC507 and UGC3355. We identify counterparts for all of these members (see the third section of \cref{membertable}) and reproduce this filamentary appearance to a striking degree. This can be seen in the top row of \cref{P-P}: The V-like structure of the main branches of the filament given by A426 (1), AWM 7 (2) and UGC2562 (3) found in the observation is reproduced by the simulated region, as indicated by the gas density contours connecting the two companions to A426.
\subsubsection{The Coma supercluster region}
The Coma supercluster or, as it is alternatively called, the Great Wall \citep{gott2005}, has 5 massive member clusters identified by \citet{boehringer2021}: A1656, A1367, A1185, MKW4 and NGC4325. We find counterparts for the first four of these clusters in the SLOW simulation.
The main halo A1656 - the Coma cluster - is one of the most prominent and best-studied structures in the local universe. The properties of the cross-matched members are listed in \cref{membertable}. Especially the region spanning triangle of the Coma cluster, MKW4 and A1185 constrains this supercluster region well as can be seen from the red dots in \cref{Mollweide} and in more detail in \cref{Comaquad} where a zoom-in view of the region is provided both in observations and in simulations as well as the collapse structure of the supercluster and its large scale (dark matter) environment. 

\subsubsection{The Local supercluster region}
The Local supercluster is a typical spider-type supercluster \citep{einasto2007a}, dominated by one central rich cluster (Virgo) which is also the only member of this region from the catalog by \citet{boehringer2021} above $10^{14}M_\odot$. We identify a match for this cluster as well as  another less massive system, NGC5846, from the catalog in the SLOW simulations. Otherwise, the Virgo supercluster is comprised by galaxies in a filamentary arrangement. This property is well reproduced by SLOW as can be seen in the lower right figure of the four panel plot \cref{Virgo} and in the three projections of \cref{Virgohal}, with the dominant filament being the Virgo-Centaurus filament. Additionally, we identify a promising counterpart for one of the "traditional" local supercluster members, namely the M94 group (sometimes referred to as CVn or CVn I group), that was omitted in the catalog by \citet{boehringer2021}.
The Local SC is the most tentative FOF association described by \citet{boehringer2021}, requiring alleviation of the criteria the authors apply: Several of the smaller groups that the authors list as members of the structure fall below the luminosity limit applied to the other superclusters and additionally they list the giant ellipticals M86 and M87 as separate members, despite belonging to the Virgo cluster itself. The problem of the relatively low mass of the members identified by \citet{boehringer2021} is further aggravated in the simulations by the uncertainty of the center i.e. the position of the observer in the model universe, see \citet{dolag2023a}. Additionally, as an extremely close by region the geometric distortions of a shift in center are expected to be the most severe for this supercluster. Correspondingly, the members of this region - as defined by \citet{boehringer2021}- have only relatively weakly constrained counterparts in the simulation or no counterpart can be reliably defined with our methods.
\subsubsection{The Centaurus supercluster region}
The Centaurus supercluster, as identified by \citet{boehringer2021} hosts only two massive structures, A3526 and NGC5044. We find a counterpart for A3526 in the SLOW simulation and additionally identify matches for three of the less massive systems, NGC4936, AS0753 and A3574E.
Whether the Hydra and Centaurus supercluster region comprises of one joint supercluster rather than two separate agglomerations is a matter of debate in the literature. \citet{boehringer2021} find these regions to be separate using their luminosity-weighted linking length. However, in the SLOW simulation, the distance of Hydra and Centaurus is reduced significantly ($r_\mathrm{sim}=9.6$ Mpc) compared to the observed distance  ($r_\mathrm{obs}=30.9 \pm 6.0$ Mpc)\footnote{Relative distance and error are based on the positions and velocity uncertainty given by \citet{tully2015}}. As a result, the two clusters are gravitationally bound together with Hydra falling onto Centaurus. Thus the supercluster regions are merged into one, consequently also biasing the masses in the region towards higher masses. The closer proximity of the Hydra cluster compared to the observed Centaurus supercluster additionally changes the geometry of the region, complicating the identification of clear counterparts for the member clusters, especially NGC5044, the most massive secondary cluster in this region.
\subsubsection{The Hercules supercluster}
\label{sssec:hercules}
\paragraph{Cross-identification of A2147}
Because \citep{hernandez-martinez2024a} do not find a counterpart for the main cluster of the Hercules supercluster structure, namely A2147, applying their search criteria, we start by trying to identify a counterpart with a slightly extended search radius. In their study on the Hercules supercluster core, \citet{monteiro-oliveira2022} give a total mass of $13.5^{+2.1}_{-1.7}\times10^{14}M_\odot$, placing it well into the mass regime of the most massive halos in the simulation box. This means that a detection, even at a higher distance to the observed position, is likely to be reliable. We identify a candidate halo at $r_{\mathrm{obs}}=53.54$ Mpc/h separation from the observed position. We compute the $M_{500}$-based significance as described by \citet{hernandez-martinez2024a} for this detected counterpart. The resulting probability is $p=0.043$, indicating a detection probability of $95.7\%$, indicating that this halo is a viable counterpart (see also \cref{A2147sig}).  

 The Hercules supercluster core itself is a concentrated structure ($r_{proj}\approx$ 9~Mpc according to observations \citep{monteiro-oliveira2022}) constituted by three main structures: A2147, A2151 (Hercules cluster), and A2152. \citet{monteiro-oliveira2022} additionally identified two massive sub-clumps not previously identified in the structure. In the SLOW simulation, A2147 as the central halo is fed by at least three filaments (see the contours in \cref{Hercules}) along which other massive groups and clusters such as A2151 and A2152 fall in. Both of these infalling structures have counterparts in the SLOW simulation. However, in the catalog by \citet{boehringer2021a} only the low-mass tail of the larger supercluster region lies within the survey volume, leading to relatively low-mass members ($10^{13}-10^{14}M_\odot$) in their study. Due to their distance to the supercluster core and the low masses, we currently do not find reliable counterparts for these structures in SLOW.
\subsection{The future of bound structures}
\label{sec:future}
In the linear picture of structure formation, initial density fluctuations of small to intermediate wavelengths enter their collapse and virialization phase first, followed by the long wavelength modes. For this reason, galaxies and even clusters of galaxies form relatively early on in a phase of moderate background cosmology expansion. The biggest structures, however, namely superclusters, are only just forming today, making the cosmological context they evolve in a quite different one to the matter-dominated background universe at $z=0.5$ and higher. Their relatively young dynamical age at the present time makes superclusters hard to separate and it is unclear to which degree structures that appear connected in observations can actually start collapsing and form virialized structures. In \cref{ssec: Clairvoyant} we introduced the additional forward propagated box of the SLOW initial conditions. \cref{Clair} shows a slice through this \texttt{Clairvoyant} simulation at the present time and in the far future. It demonstrates the evolution of the cosmic web to be already almost progressed to its final morphological state at the present time. It should be noted here that this standstill occurs only in co-moving space, in real space the frozen structure is rapidly drifting apart.

There is little to no late time evolution in co-moving space at the largest scales, consistent with previous numerical experiments \citep[e.g. ][]{nagamine2003a,busha2003,hoffman2007,araya-melo2009a}. Collapsing overdensities separate causally into gravitational "islands" with all flows between those regions halted by the rapid expansion pulling theses islands apart in real space. For similar reasons, while the voids continue to empty out after the present time, there is also a significant amount of matter that is left behind in the voids, never reaching the walls on its outflowing trajectory. 

\begin{figure*}[ht]
   \centering
   \includegraphics[width=\textwidth]{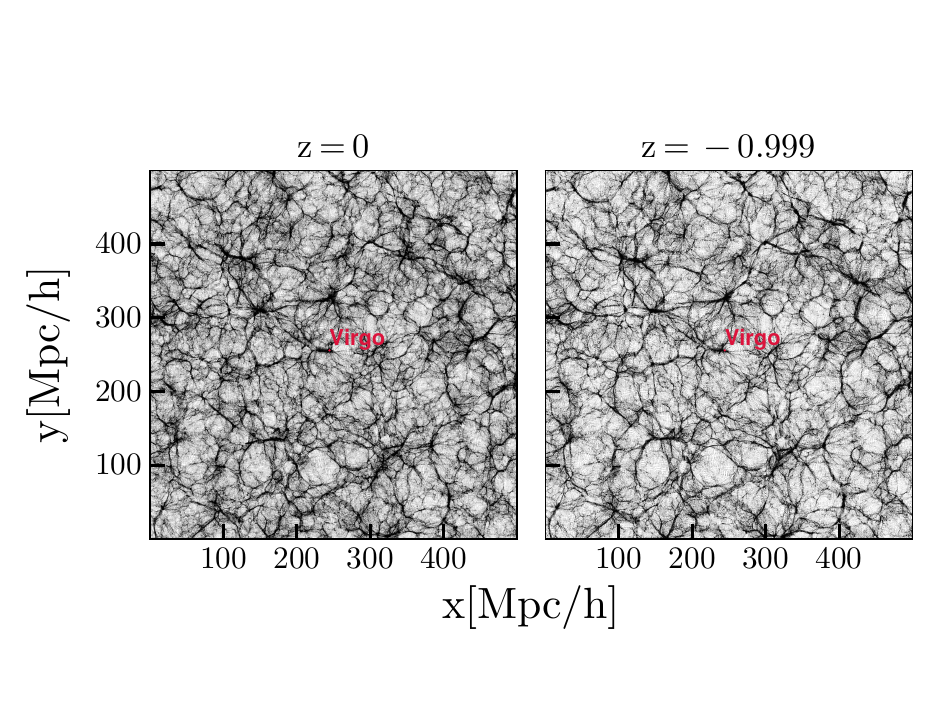}
   \caption{The cosmic web in the \texttt{Clairvoyant} simulation (dark matter component) within a slice of 10 Mpc/h at the present time (left) and in the far future (right). The position of the Virgo cluster as the nearest cluster to the observer position is marked in crimson.}
              \label{Clair}%
    \end{figure*}
    The suppression of structure formation is also reflected in the evolution of halo masses. \cref{hmf} shows the halo mass  in the Clairvoyant simulation and its evolution with cosmic time. While the halo mass function (hmf) evolves to some degree after $z=0$, barely reaching objects with masses of $10^{16}M_\odot$, its evolution is severely slowed down. Additionally, it is precisely the largest structures that assemble after the present time, with a completely halted evolution at the group to small cluster mass end. This reflects the assembly of the supercluster structures that are dynamically still very young at the present time. Beyond 60 Gyr the evolution of large-scale structures is stopped completely in co-moving space as can be seen by the last halo mass function lines completely overlapping at the high-mass end. 
    \begin{figure}[ht]
   \centering
   \includegraphics[width=\columnwidth]{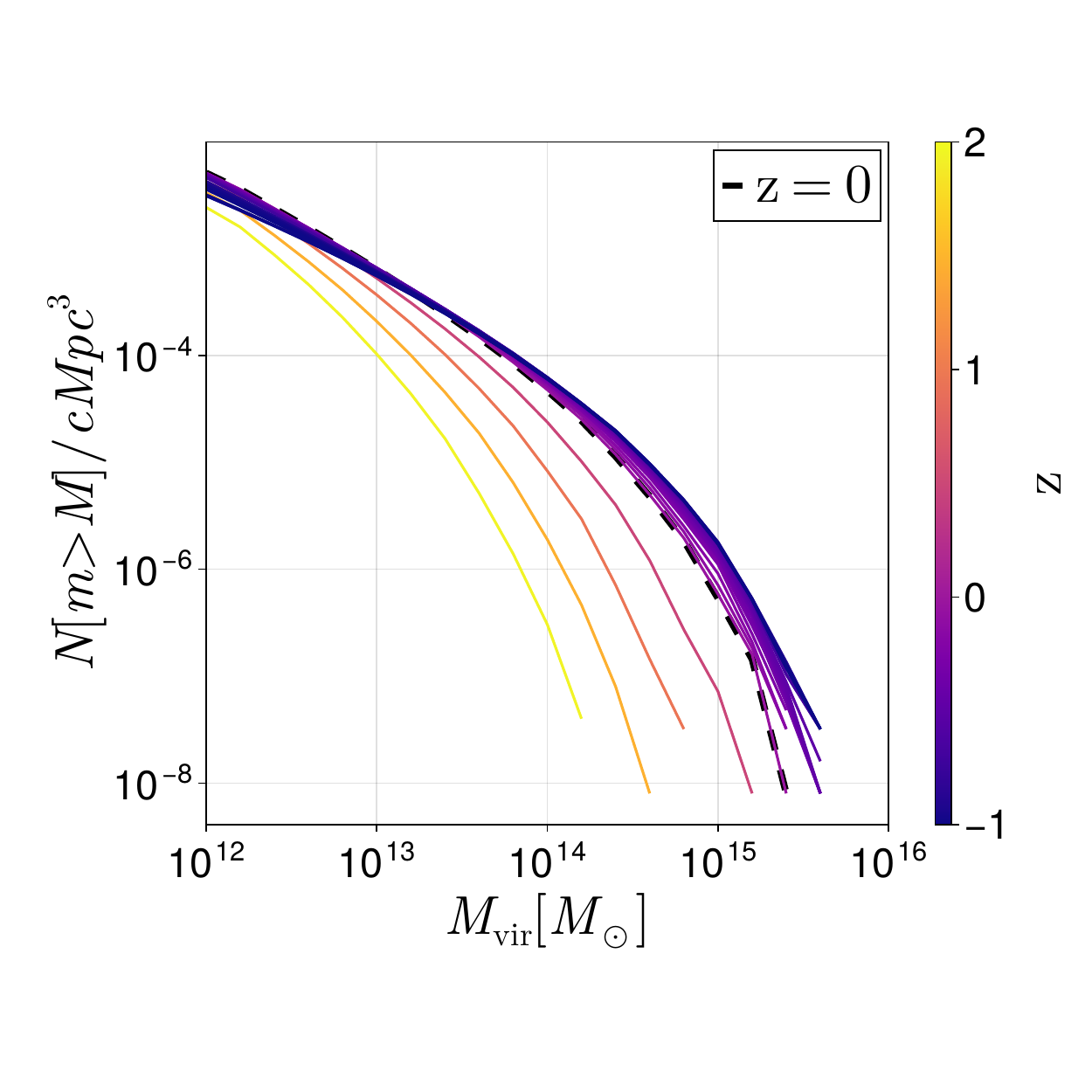}
   \caption{Evolution of the local halo mass function far into the future. Each line represents the hmf at the redshift indicated by the line color The present time hmf is the dashed purple line. }
              \label{hmf}
    \end{figure}
    \begin{figure}
    \centering
    \includegraphics[width=\columnwidth]{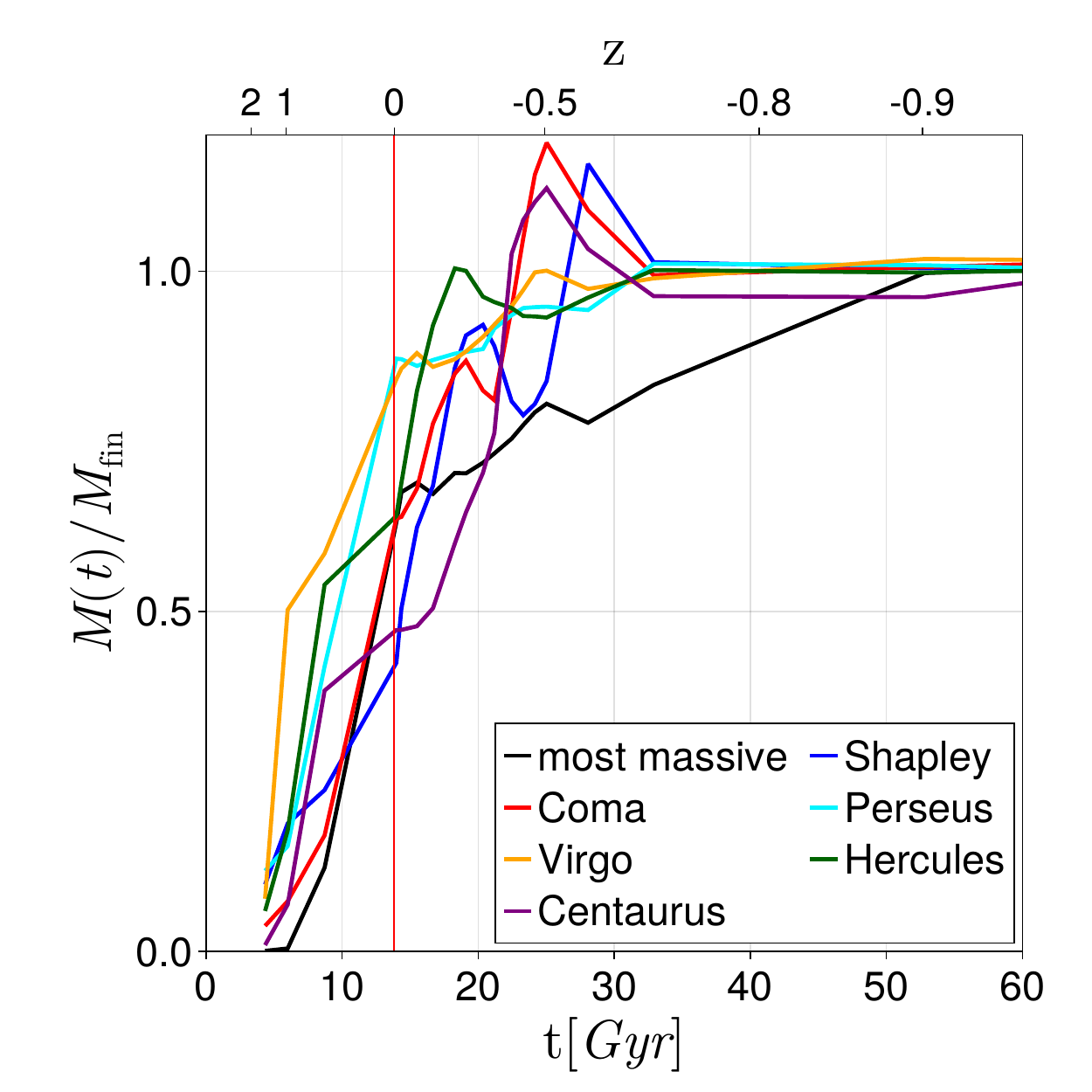}
    \caption{Late-time accretion histories for the six supercluster regions introduced in the previous section relative to their respective final halo masses.}
    \label{sampletrace}
    \end{figure}
    
    \cref{sampletrace} shows the mass evolution of the haloes associated with the superclusters introduced in the previous section in the \texttt{Clairvoyant} simulation. We furthermore show the evolution of the most massive halo in the simulation volume overall (black line) as a reference for the general structure evolution in the box. Generally, two classes of superclusters can be identified w.r.t. their mass growth history: Those that have already assembled their mass fully at the present time (Perseus and Virgo) and superclusters that undergo significant mass accretion and major mergers in the future before freezing out. 
    
    The most prominent example of such late growing superclusters is Shapley: This supercluster experiences two major mergers in the future, associated with the infalls of A3560 and A3571 respectively. A common peculiarity of these late-time mergers is that the virial mass of the halos actually decreases again after reaching a peak after the secondary cluster has fallen in. This peculiarity can also be understood with the late time merger of the Shapley supercluster with A3571 (the rightmost spike of the blue curve in \cref{sampletrace}):
    While A3571 reaches the cluster center against the expansion of the background and can thus be considered bound to the survivor cluster, a significant amount of its mass is lost again by the subsequent puffing up of the halo. This leads to particles being ejected to radii where they can rejoin the Hubble flow, keeping them from collapsing to the final virialized object and subsequently decreasing the mass within a virial radius of the halo again (although in total the mass increases). 
    
    The effect of post merger-relaxation has been studied by \citet{lee2018}, albeit at z=0. They find that significant post-merger mass losses are common, with low-mass halos being more strongly affected by this effect than high-mass halos. Interestingly at z=0 in their highest mass bin centered at $\mathrm{log}(M[M_\odot])=13.45$ they find a fraction of $\approx 5\%$ of halos that experience a mass-loss greater than $10\%$ of the peak mass (i.e. the maximum mass the halo reaches in the merging process). In our (statistically not significant) sample, $50\%$ of halos lose more than $10\%$ of their peak mass, with Coma achieving the highest loss fraction of $\approx 16\%$ (see also \cref{Sups}). Considering the mass trend found by \citet{lee2018} and the masses of the halo sample in this work, this is a strong indication that this mass-loss process is much more drastic in the late-time universe, aided by the rapid expansion. Another distinction to the mass loss studied by \citet{lee2018} is that for the late-time mergers the mass loss is generally not recovered through re-accretion of the lost material. This is illustrated by the trajectory of Hercules in \cref{sampletrace}: Due to the final major merger happening early enough ($t\approx 18Gyr$), the supercluster is able to recover nearly all of its peak mass after an initial post-merger mass loss. In contrast, the mass lost by the last major mergers in e.g. Coma is lost to the Hubble flow and never recovered.

    The concept of "unbound" particles was previously discussed in the literature by e.g. \citet{behroozi2013}, though this study was limited to massive haloes that do not undergo a major merger after z=0. Nevertheless the finding that simple energy considerations are \textit{poor} predictors of whether particles are actually bound to a given structure is in excellent agreement with the diversity of mass history trajectories and irregular collapse volumes we find in this study (see e.g. \cref{Shapley} and the more detailed discussion in \cref{collapse}).  

    \subsubsection{Freeze out time}
    To more accurately estimate the freeze-out time, the time at which structure formation is stopped completely, we consider the mass assembly histories of the most massive structures, namely the most massive supercluster from the sample, the Shapley supercluster, and the halo with the largest final mass in the box overall (black line) shown in \cref{trace}. Though the overall mass growth is already suppressed significantly after $\approx 2 t_H$, there is still considerable growth in mass for both halos after this point, finally evening out at a freeze out time of $t \approx 50$ Gyr (the most massive halo in the simulation evens out its growth to a growth rate of less than $0.1\%$ of its final mass ($M_\mathrm{vir,fin}=5.5\times10^{15}M_\odot$) per Gyr at 52.85 Gyr). Interestingly, compared to the simulation run by \citet{nagamine2003a}, which used a similar approach and found a freeze-out time of about 30 Gyr in the future (or $\approx t_0+2t_H$ with their parameters) this convergence time is further in the future. \citet{nagamine2003a} used different cosmological parameters ($\Omega_0=0.3,\Omega_\Lambda=0.7,h=0.7$) and a smaller simulation volume with vacuum boundary conditions. All of these factors can impact the exact timing of the freeze-out: The slightly higher dark energy density and larger $H_0$ used by \citep{nagamine2003a} shift the freeze out closer to z=0 because the background expansion dominates against gravity at an earlier time. A sensitivity to the $\sigma_8$ used for the simulation is also likely as the strength of the clustering can additionally affect the exact collapse times, unfortunately \citet{nagamine2003a} do not provide this normalization in their work.
    
    One could expect the vacuum boundary conditions to result in more strongly bound objects and possibly a more rapid evolution of the bound structures before the freeze-out compared to the periodic boundary conditions employed in SLOW because of the lack of large-scale tidal forces. On the other hand a larger box contains longer wavelength density perturbations compared to a smaller volume. As the superclusters, being the largest bound structures in the volume, are expected to sit at the peaks of these large-scale fluctuations, one would expect a constructive effect on the evolution of these regions in the larger box of SLOW and thus a shortening of the freeze-out time. However without a dedicated experiment, the relative importance of these effects is difficult to assess. We leave this experiment for future work as it is beyond the scope of this first study on the SLOW superclusters.
    \begin{figure*}[ht]
   \centering
   \includegraphics[width=\columnwidth]{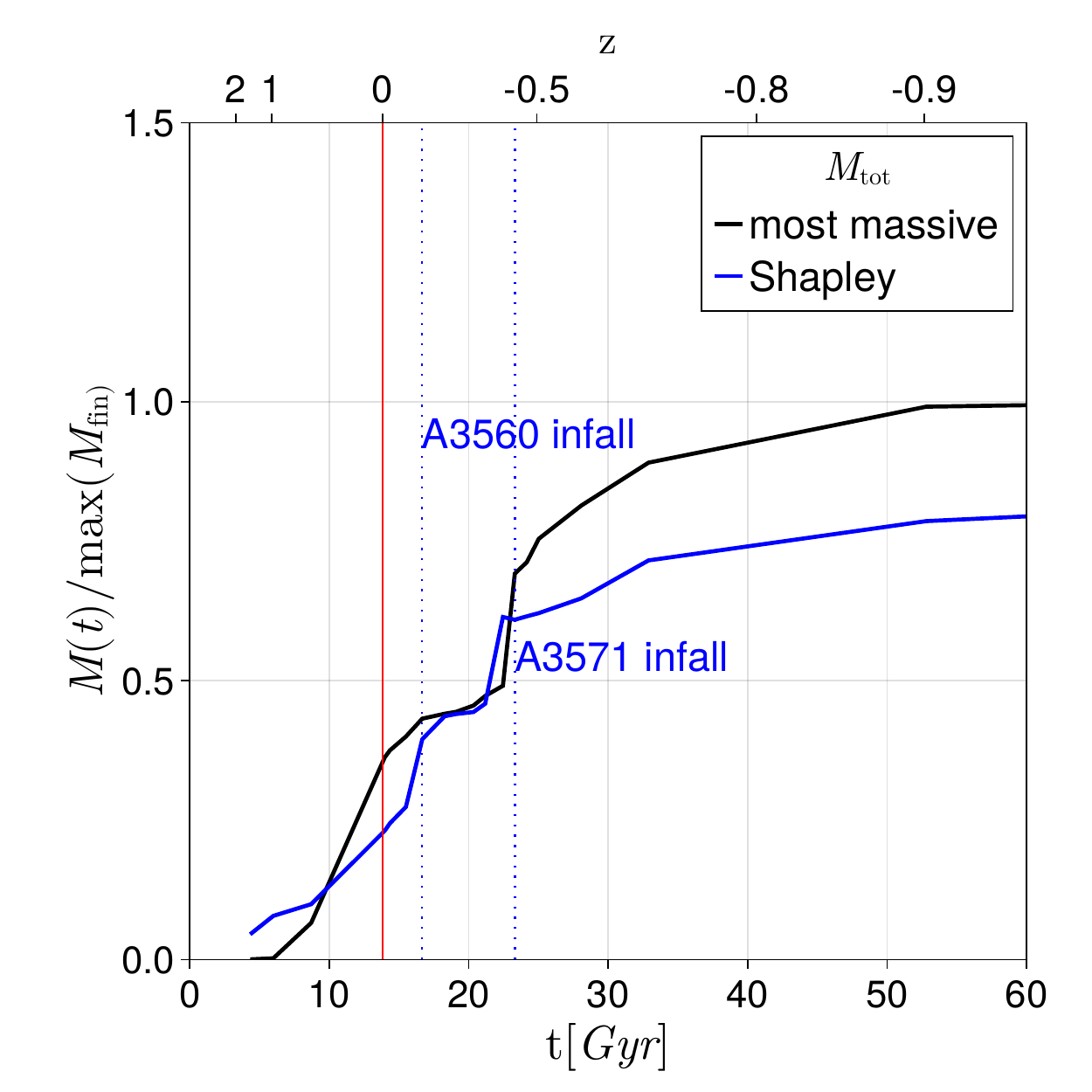}
   \includegraphics[width=\columnwidth]{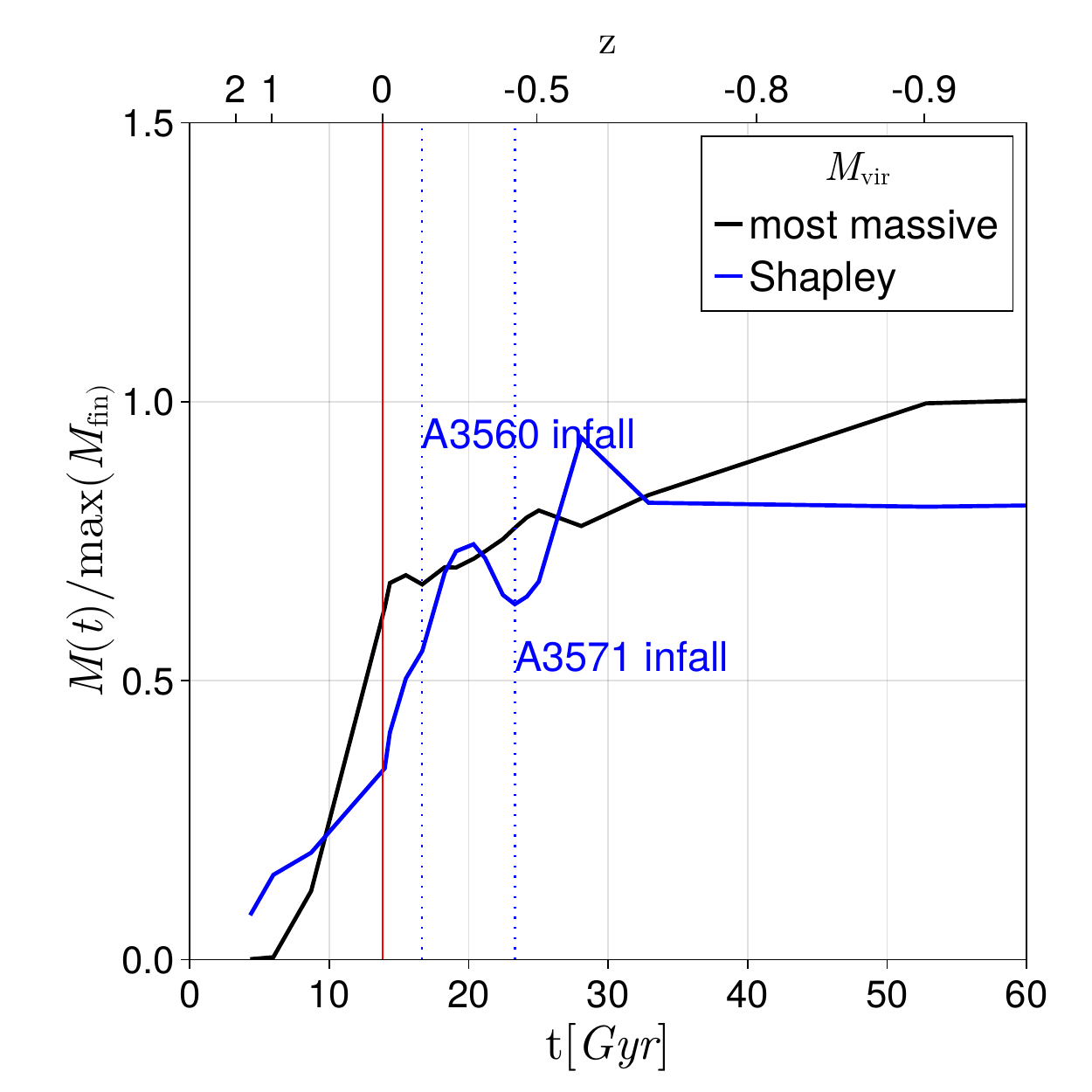}
   \caption{Mass assembly co-evolution of the most massive node and the Shapley supercluster in total FOF mass (left panel) and the mass constrained to $r_\mathrm{vir}$ (right panel). Both panels show the mass of each halo at a given time normalized to the final mass of the most massive halo in the simulation. }
              \label{trace}
    \end{figure*}
\subsubsection{Does SLOW predict the Milky Way to fall onto Virgo?}
At the present time the Local Group has a measurable peculiar velocity w.r.t. the Hubble flow e.g. \citep[e.g.][]{lynden-bell1988} into the direction of the Local supercluster. This raises the question whether the Local Group itself might be bound to the Local supercluster. Previous numerical works e.g.  tackled this question from both a numerical \citep{nagamine2003a} and analytical \citep{chon2015} perspective. All current evidence points towards the Local Group receding away from the Local supercluster in the future due to cosmic expansion. We can verify this result in the SLOW simulation by using the collapse volume of the Local supercluster from the \texttt{Clairvoyant} simulation. Since there is no direct counterpart for the Local Group in SLOW, we use the box center as the first proxy for its position. This choice is reasonable, since the constraining data are by construction centered on our position in the universe. A second plausible choice would be the SLOW-optimal center given by \citet{dolag2023a} as the Milky Way position taking into account the distortions that affect the positions of local galaxy clusters. \cref{Virgoinfall} shows both of these proxies and their position relative to the collapse volume of the Local Supercluster. In agreement with previous works, the SLOW simulation predicts the Local Group to be unbound from its closest supercluster as the collapse volume of the Local Supercluster does not contain either choice for the Milky Way/Local Group position. This also indicates the collapse volumes computed in this work to generally agree with analytical expectations, which we will explore further in the next subsection.   
        \begin{figure}[ht]
   \centering
   \includegraphics[width=\columnwidth]{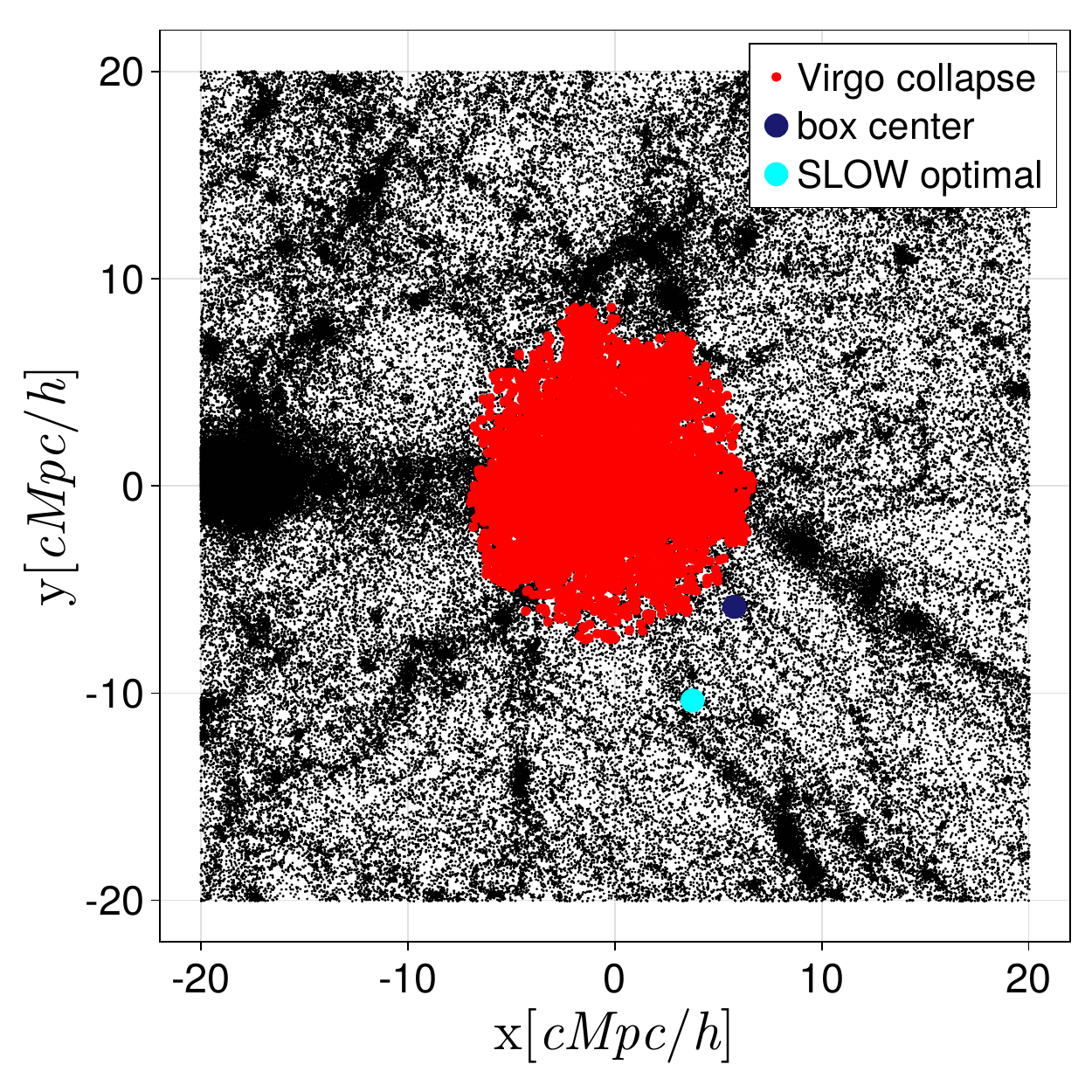}
   \caption{Collapse region of the Virgo cluster (red) and the relative position of the SLOW-Box center and the optimal center from \citet{dolag2023a} as proxies for the Milky Way position.}
              \label{Virgoinfall}
    \end{figure}
\subsection{Collapse regions}
\label{collapse}
What constitutes a supercluster region from a physical perspective is an inherently ambiguous question. The definitions from previous works can be summarized into the following three criteria:
\begin{enumerate}
    \item \textbf{Density criteria} i.e. selecting a region enclosed by some density contour e.g. \citet{boehringer2021,boehringer2021a}
    \item \textbf{Dynamical criteria} i.e. selecting mass tracers converging onto a local attractor e.g. \citet{dupuy2019}
    \item \textbf{Boundness criteria} i.e. selecting a region that will collapse onto a structure in the future respecting the background evolution of the cosmology e.g. \citet{araya-melo2009a,dunner2006,chon2015}
\end{enumerate}
There have been a number of theoretical and numerical approaches towards determining the collapsing or bound fraction of a given overdense region based on the evolution of a spherical overdensity with regards to a homogeneous background cosmology \citep{nagamine2003a,busha2003,dunner2006,hoffman2007,araya-melo2009a,pearson2013,chon2015}. These works all aimed at deriving a critical density (or mass given an observed radius) a region needs to be able to form a single structure in the future. The derivation by \citet{busha2003} arrived at a critical density ratio of $\Omega_c=5.56$ for a marginally collapsing shell at $a \rightarrow \infty$. In contrast, by relaxing the assumption of particles moving with the Hubble flow at the present time and working purely with spherical collapse theory \citet{dunner2006} obtained a lower critical density ratio of $\Omega_c=2.36$, a value later confirmed by \citet{chon2015} to be stable across different cosmologies. \citet{chon2015} referred to those collapsing regions as "superstes" clusters (lat. survivor) because they are able to survive the accelerated expansion of the universe at late times and in the future. These objects are systematically smaller and contain less mass than the supercluster regions obtained with criteria 1 and 2 and rather represent the dense cores of these associations, see \citep{chon2015} whereas especially the dynamically defined "basins of attraction" - i.e. the regions from which all large-scale flow lines converge to a single point - have been shown to partition a given cosmological volume completely \citep{dupuy2019}.

In numerical simulations, criterion 3 is accessible by letting the gravitational part of the simulation propagate far into the future and keeping track of which parts of an associated region reach the center at $t_{\mathrm{final}}$. A first-order approach to this is simply evaluating the FOF groups in the final snapshot, as described in \cref{ssec:SLOWsups}. This method associates every particle that comes within a linking length of a supercluster member particle with the selected supercluster. It will conservatively include every particle that reaches the supercluster center, which has some technical benefits for creating a new class of zoom-in initial conditions for galaxy cluster and supercluster simulations further discussed in a companion paper to this work (Seidel, Dolag and Sorce in prep.).
For a direct comparison between the method used in this work and purely theoretical approaches, \citet{chon2015} calculate the extent of the collapsing region (maximum radius to which the overdensity necessary for collapse is fulfilled) of the Shapley supercluster at z=0. \cref{Superstes} shows the comparison between this radius and the convex hull of all FOF particles from \texttt{Clairvoyant} for this supercluster. It can be seen that the total extent is in good agreement, with deviations arising mainly from the more adaptive shape of the Lagrangian region.
    \begin{figure}[ht]
   \centering
   \includegraphics[width=\columnwidth]{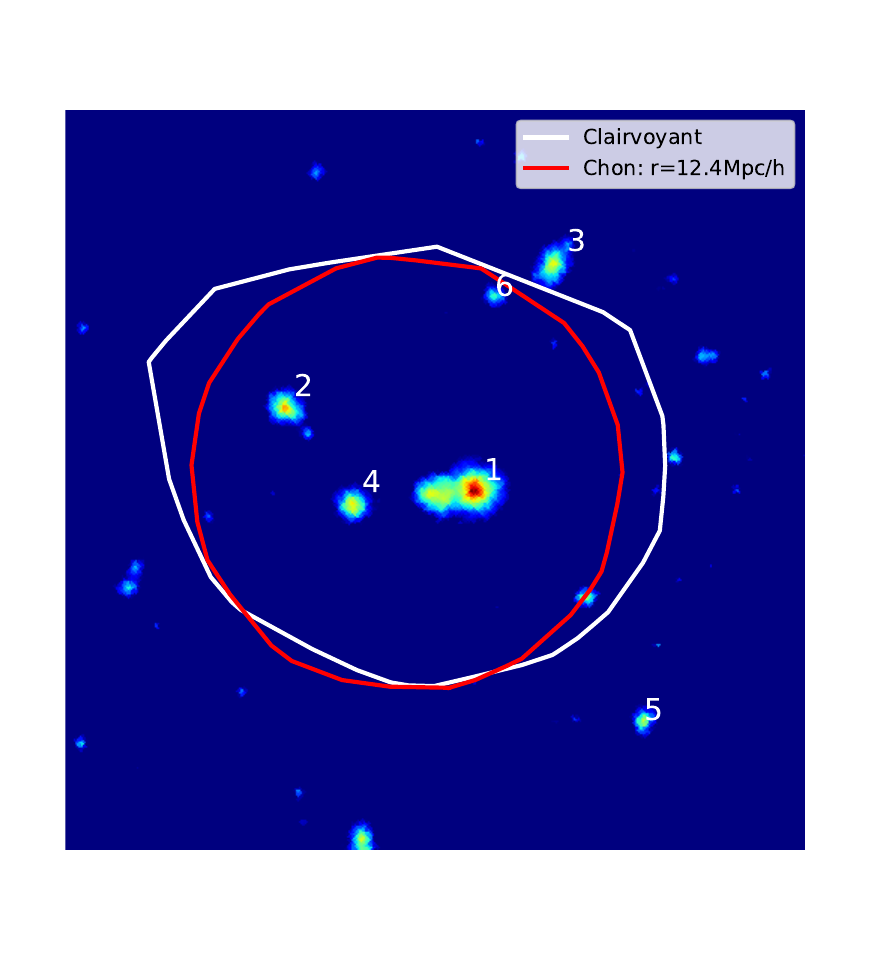}
   \caption{Mock observation of the thermal Sunyaev-Zeldovich signal indicating the shape of the Shapley supercluster simulated collapse region at z=0 (white) vs. predicted extent of the "superstes" region (red) today given by the bound radius derived by  \citet{chon2015}. The number labels are identical to \cref{Shapley}. 1:A3558, 2:A3571, 3:A1736, 4: A3560, 5:A3532, 6:A3559 }
              \label{Superstes}%
    \end{figure}
\begin{table}[h!]
\tiny
\begin{tabular}{lllllll}
Region& $N_{\mathrm{m}}$&$M_{\mathrm{FoF}}$&$M_{\mathrm{vir}}$&$M_\mathrm{est}$ (reference)&$\frac{M(z=0)}{M_\mathrm{fin}}$& $\frac{M_\mathrm{fin}}{M_\mathrm{peak}}$\\
\hline
Shapley        & 3         &   72.86              &    44.42& \makecell[l]{500.0\\ \citep{proust2006}}&0.42&0.86\\
Perseus-Pisces & 2         &  18.09                &  12.13&249.0 (B2)&0.87&0.99\\
Coma           & 2         &  47.08                &  29.52&168.0 (B1)&0.64&0.84\\
Virgo          & 2         &  16.59                 & 11.91&25.0 (B1)&0.84&0.98\\
Centaurus      & 2         &    34.56             &   21.59& 56.0 (B1)&0.47&0.89\\
Hercules & 3&38.13&24.05 &21 $\pm$ 2 (MO)&0.64&0.99
\end{tabular}
\caption{Survivor Haloes of the local superclusters, the number of accreted members (out of the previously listed supercluster members) and their masses at the final snapshot ($a=1000$). All masses are given in units of $10^{14}M_\odot$.}
\label{Sups}
\end{table}
\cref{Sups} lists the final collapse masses of the introduced regions. It is apparent that conventional linking methods such as the ones used to link the superclusters in the catalogs B1,B2 and \citep{proust2006} systematically over predict the mass that is actually bound to these superclusters. This usually only contains the central most companion halos as can be seen in \cref{membertable}. The exception from this in our sample is the Hercules supercluster, where the mass estimate from the observation is based only on the innermost core members, which are gravitationally bound to the central halo due to their proximity at z=0 according to \texttt{Clairvoyant}. Notable is also that the two most massive regions in the local supercluster set have the lowest ratios of FOF mass to mass contained within the virial volume. This fact is related to the mass loss these halos experience when they undergo a massive merger, as discussed earlier. The lost particles stay within the FOF density contour but remain frozen outside of the virial surface due to the background expansion.

One of the predictions made by spherical collapse models like \citet{dunner2006} is the \textbf{Critical density contrast} of marginally collapsing regions at z=0:
$\Omega_{\mathrm{col}}=\frac{\rho_\mathrm{col}}{\rho_\mathrm{crit}}$. Numerically we compute this ratio for the collapsing supercluster regions ($\Omega_{\mathrm{SC}}$) from the forward simulation via 
\begin{equation}
\Omega_{\mathrm{SC}}=\frac{\rho_\mathrm{SC}}{\rho_\mathrm{crit}}=\frac{M_\mathrm{SC}}{\rho_\mathrm{crit}V_\mathrm{SC,z=0}}
\end{equation}
where $M_\mathrm{SC}$ is the mass of all particles in the previously defined collapse regions and $V_\mathrm{SC,z=0}$ is the Volume occupied by them at z=0 estimated by computing their 3D convex hull. 
\cref{denscon} shows how the present-time density contrast of these particle groups compares to the theoretical predictions from spherical collapse theory \citep{dunner2006}. It is apparent that, while the theoretical prediction is a good estimate, the scatter of the numerically obtained densities of the collapse volumes is quite high. Because the spherical estimate from \citet{dunner2006} considers only radial motions, there is a straightforward explanation for the clusters exhibiting a higher density ratio than predicted: These halos develop in different tidal environments and, due to these large scale tidal fields, accrete angular momentum differently. Since angular momentum hinders the gravitational collapse, a higher density contrast is needed for the region to be bound. This argument is similar to the one used by \citet{dunner2006} to explain the difference to the numerical experiments performed by \citet{busha2003}, which yielded a significantly higher critical density.  
\begin{figure}[ht]
   \centering
   \includegraphics[width=\columnwidth]{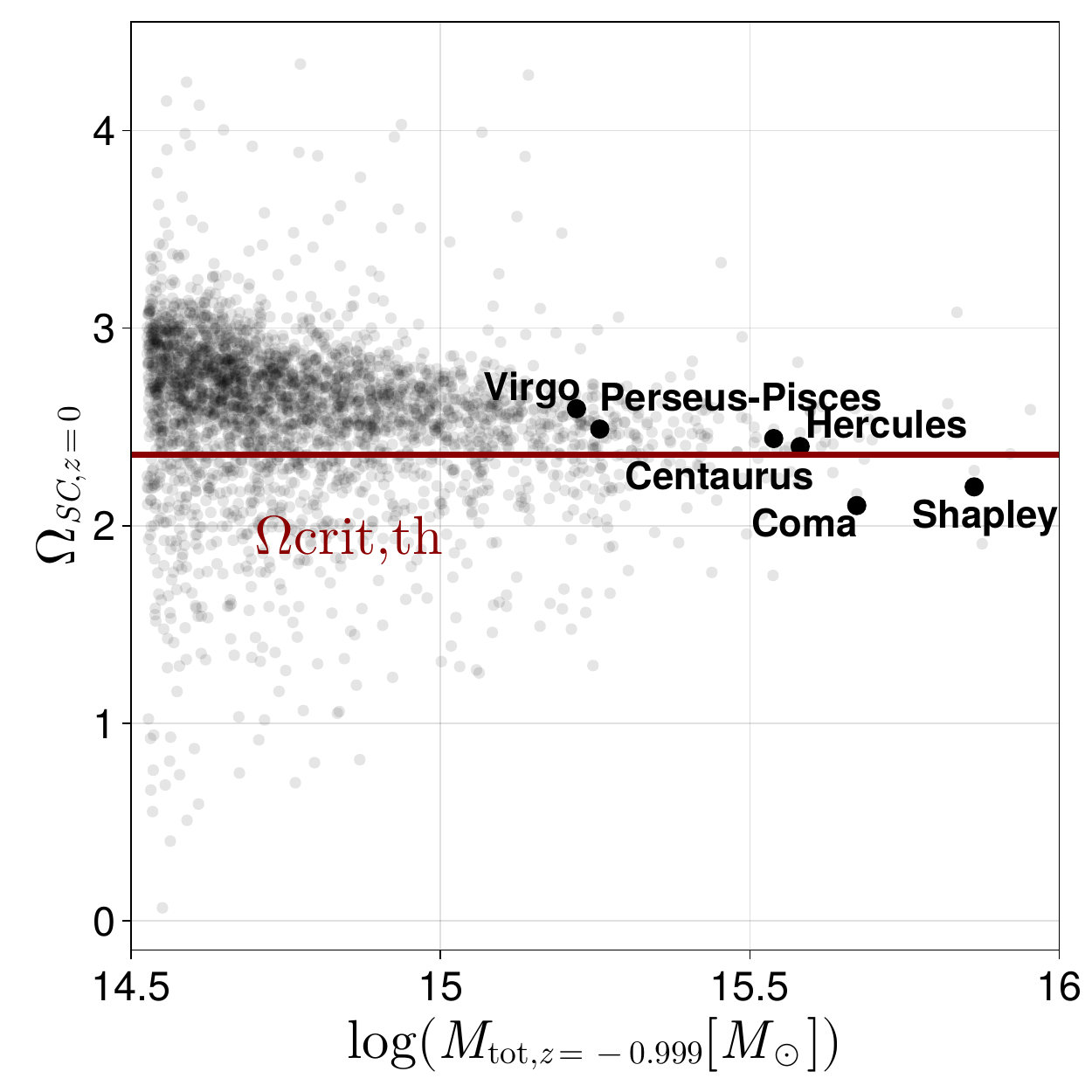}
   \caption{Density ratio at $z=0$ of the final FOF groups of the supercluster regions identified in the \texttt{Clairvoyant} simulation compared to the critical collapse density contrast derived by \citet{dunner2006} (red horizontal line). The 2000 most massive halos at z=-0.999, corresponding to the most massive superstes volumes, are shown additionally as transparent dots.}
              \label{denscon}
    \end{figure}
As can be seen from \cref{trace} and from the masses in \cref{Sups}, even particles that make it into the final FOF group can actually end up and freeze (in comoving space) \textit{outside} the halo by virtue of the finite timescale of virialization. A prerequisite for this complication is a very late infall as demonstrated by the infall of A3571, as discussed before. An overall consequence of this is a more diffuse final dark matter halo as is evident from the density contrasts in \cref{denscon} for the Shapley and Coma supercluster regions. 
\section{Summary and Conclusions}
We used the SLOW constrained simulations of the local universe to study nearby superclusters. By matching these structures and their member clusters pairwise to their observed counterparts we are able to compare basic properties of these clusters, such as their masses, their relative positions and their X-ray luminosities to observations. By running an N-Body version of SLOW, the \texttt{Clairvoyant} suite into the far future (z=-0.999), we are furthermore able to determine the final masses of these superclusters and discern which member clusters are actually bound to them. We furthermore compare these results to previous numerical and theoretical estimates. Our main conclusions are:
   \begin{enumerate}
      \item In total we cross-identify supercluster members in 6 observationally identified regions in the SLOW simulation with their observed counterparts. Most of the supercluster members with masses exceeding $10^{14}$ can be reliably recovered in the simulation, providing a number of viable targets for future studies and expanding on the existing catalog of cross-identified galaxy clusters in the SLOW simulation from \citet{hernandez-martinez2024a}. Examples for such future targets are the Shapley supercluster core (A3558, A3571, A3560 (and A3562) where the planned high resolution zoom-in simulations based on the collapse volumes introduced in this work can provide insightful comparisons and context for the rich multi-wavelength observational data for this region. 
      \item We introduce a quantitative method to estimate how likely a given cross-matched galaxy cluster pair is to be a result of the constraints rather than random clustering from the simulation. According to this criterion, only few of these members can be confused with random cluster pair arrangements, with the majority of ($77 \%$ of secondaries) cross-identifications exceeding $80\%$ significance. This shows that the SLOW simulation not only reproduces the massive clusters within supercluster regions with unprecedented accuracy, but additionally also mimics their three dimensional geometric arrangement by providing suitable structures close to the positions where one would expect them based on the observed 3D arrangement of the member clusters within the supercluster. This builds on previous works concerning the environments and collapse dynamics of local galaxy clusters e.g. \citet{sorce2024, malavasi2023} and will contribute to future studies of the detailed properties of the individual member clusters e.g. their turbulent pressure support and non-thermal and thermal emission features, as the environments these members evolve in are well constrained. 
      \item The Clairvoyant forward N-Body integration of the SLOW initial conditions shows the local dominant galaxy clusters to have largely finished their evolution by $t=33$ Gyr, with the less massive regions like Perseus and Virgo reaching their final gravitational state effectively at $z=0$. 
      \item A notable feature of the late-time evolution is the massive mergers some of these clusters undergo as late as $t=2t_H$. In this process these structures undergo as much as a $16\%$ loss of their peak mass. This is a significantly higher fraction of merger mass loss than previous work found for z=0 \citep{lee2018}. At late times in the $\Lambda$ dominated universe, these mass loss processes appear to be enhanced. Furthermore for these very late time mergers the peak mass is never reached again due to the particles freezing out in co-moving space (or accelerating away with the Hubble flow in real space).
      \item In general the Clairvoyant volume completely stops its mass evolution approximately 40 Gyrs into the future when the most massive node in the simulation stops growing. Compared to previous results \citep{nagamine2003a,hoffman2007} this is late for a large-scale freeze-out but can most likely be attributed to small changes in the assumed cosmology and numerical modeling significantly impacting the exact timing of these late-time dynamics.
      \item Determining the density ratios of the Lagrangian collapse regions of massive clusters at $z=0$ shows good agreement with predictions from spherical collapse theory with the scatter explained by the different accretion histories these objects undergo after the present epoch. 
   \end{enumerate}
Building on these results in future work we plan to study the supercluster regions introduced above in greater detail, additionally making use of zoomed-in initial conditions built on the Lagrangian collapse regions shown above (Seidel, Dolag and Sorce in prep.). Furthermore we plan to identify supercluster regions in the SLOW simulations with a blind search in order to compare these superclusters with the set introduced in this work.
\begin{acknowledgements}
This work was supported by the grant agreements ANR-21-CE31-0019 / 490702358 from the French Agence Nationale de la Recherche / DFG for the LOCALIZATION project. NA acknowledges support from the European Union’s Horizon 2020 research and innovation program grant agreement ERC-2015-AdG 695561. KD acknowledges support by the Excellence Cluster ORIGINS which is funded by the Deutsche Forschungsgemeinschaft (DFG, German Research Foundation) under Germany’s Excellence  Strategy – EXC-2094 – 390783311. BS, IK and KD acknowledge funding for the COMPLEX project from the European Research Council (ERC) under the European Union’s Horizon 2020 research and innovation program grant agreement ERC-2019-AdG 882679. The calculations for the simulations were carried out at the Leibniz Supercomputer Center (LRZ) under the project pn68na (SLOW). We acknowledge support from the Center of Advanced Studies (CAS) of the LMU through the Research Group "Complex Views and New Clues of the Universe".
\end{acknowledgements}
\bibliographystyle{aa} 
\bibliography{Scbbib.bib}
\appendix
\section{Selecting from the local halo distribution.}
   \begin{figure}[ht]
   \centering
   \includegraphics[width=\columnwidth]{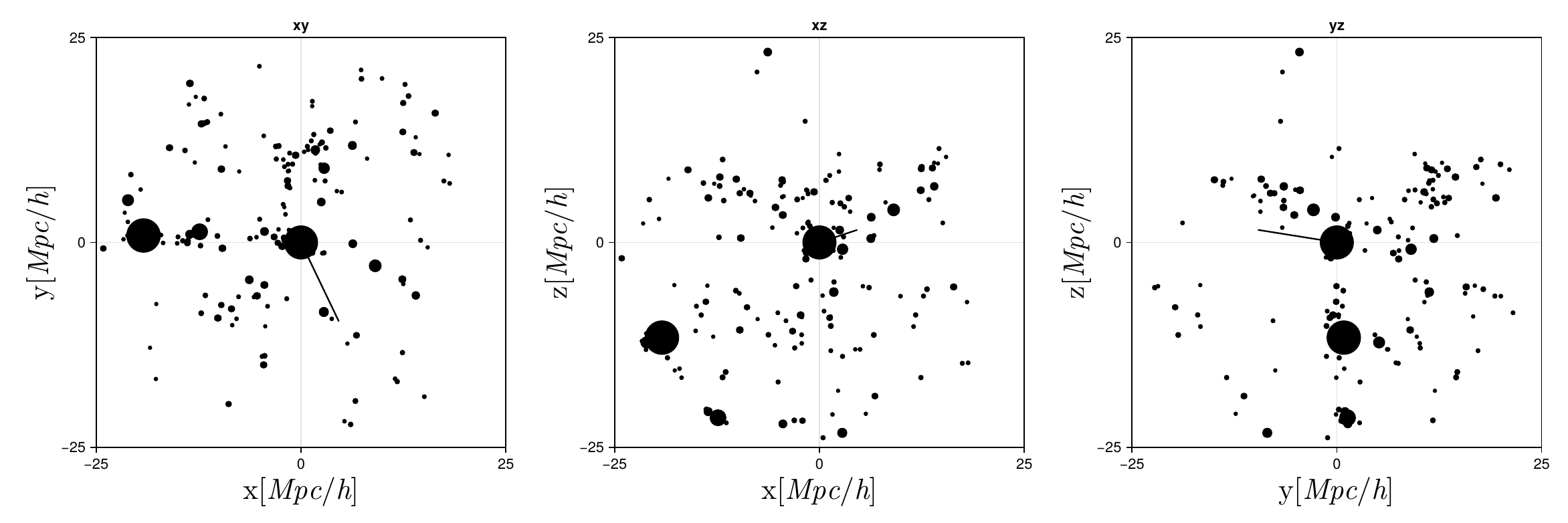}
   \caption{SLOW haloes in the vicinity of the Virgo (center) counterpart in three different projections. The point sizes are scaled with the square root of the virial masses. The black line indicates the vector separating the observed counterpart of the secondary halo to be matched (here NGC5846) and the observed primary (the Virgo cluster).}
              \label{Virgohal}
    \end{figure}
Starting from the main clusters of supercluster regions, which are often well reproduced in their simulation due to their mass and richness, we use the structure of supercluster regions to identify the more unconstrained secondaries. To this end we first take the observed positions of the main and the secondary to match in super-galactic coordinates and calculate $r_\mathrm{obs}$, as it is also used in the two-point significance approach described in \cref{sec:significance}. This gives an indication for the rough relative position of the desired halo, based on which we then select halo candidates and pick the best matches in terms of mass, angular separation and radial separation with regards to the observed relative position. A visualization of this process is shown in \cref{Virgohal}.

\section{Supercluster mock observations, collapse regions and two-point significances.}
\label{AppendixB}
In addition to the regions shown in \cref{ssec:SLOWsups} we give the same type of multiplot for the remaining regions. \cref{Comaquad} shows the very well reproduced Coma region, with Virgo in the foreground. \cref{Centaurus} shows the Centaurus region with the Hydra counterpart that does not belong to the supercluster set according to the observations. In \cref{Virgo} the Virgo supercluster can be seen with a filament connecting Virgo to the M94 group. Finally, \cref{Hercules} shows A2147 with the two nearby clusters A2151 and A2152, where the gas density contours show the Hercules cluster A2151 to be isolated from the central structure while A2152 is directly connected to it, in agreement with \citet{monteiro-oliveira2022}. Nevertheless both companion clusters are bound to A2147.
\begin{figure}[ht]
   \centering
   \includegraphics[width=0.5\textwidth]{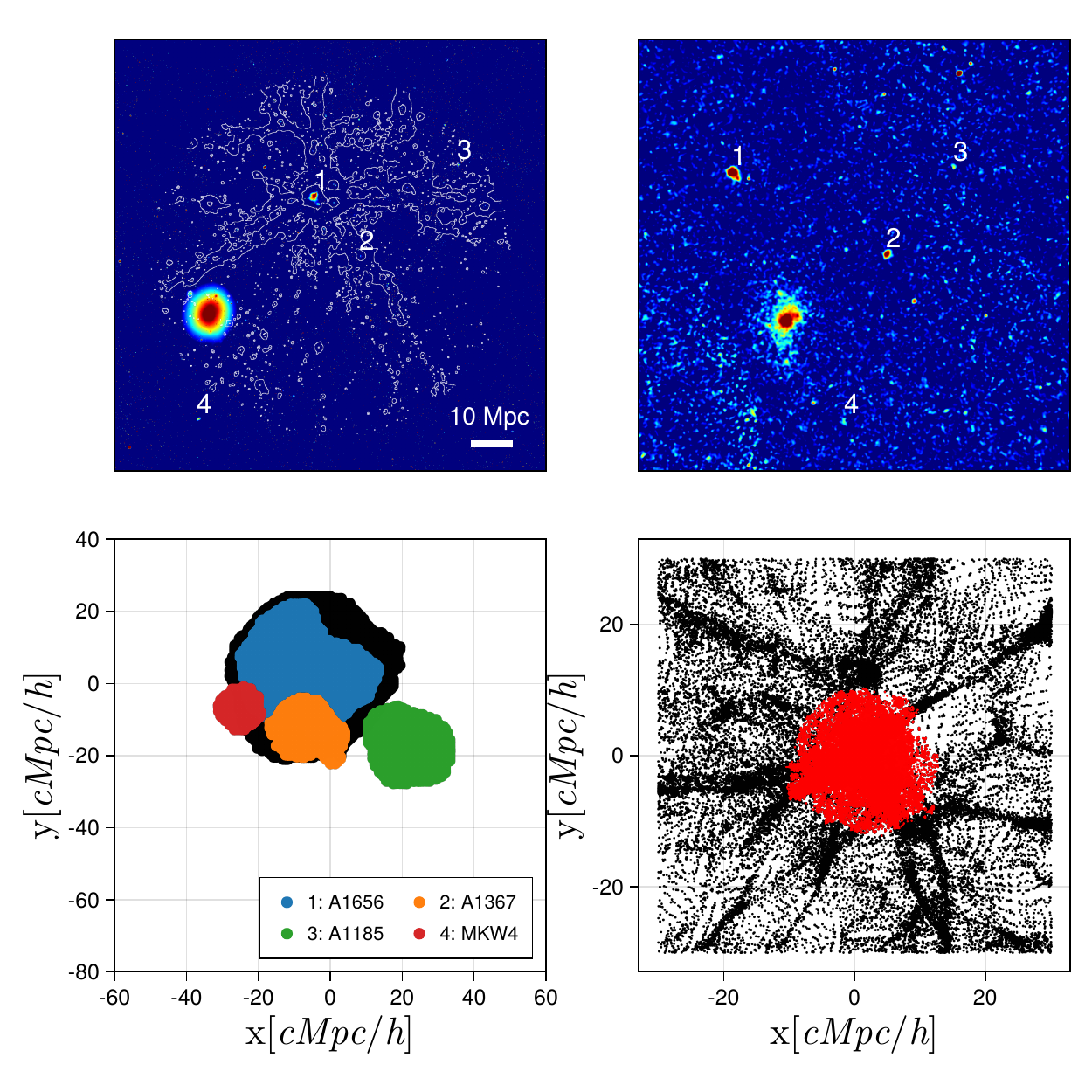}
   \caption{The Coma supercluster region in the SLOW simulation compared to ROSAT.}
              \label{Comaquad}
\end{figure}
    
    \begin{figure}[ht]
   \centering
   \includegraphics[width=0.5\textwidth]{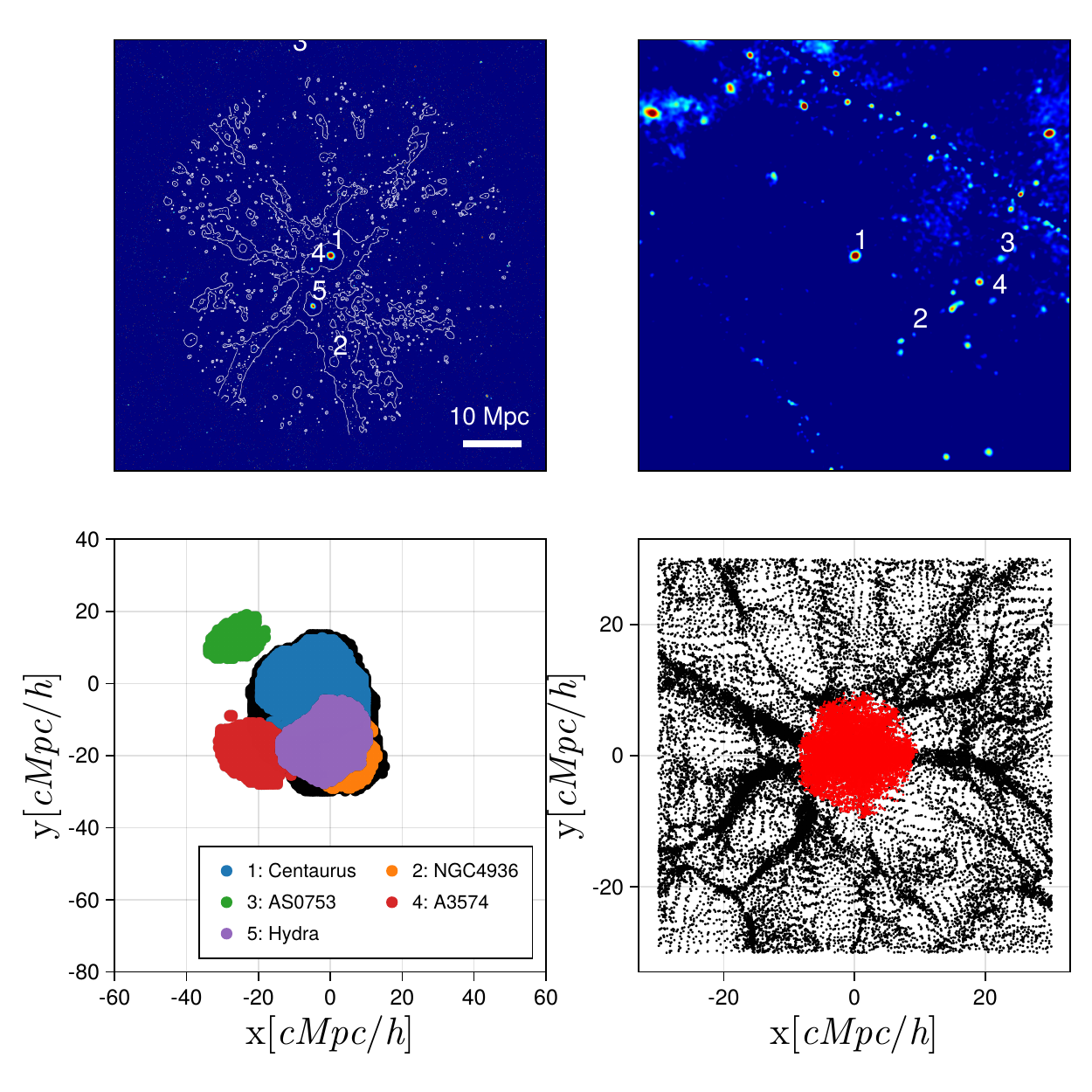}
   \caption{The Centaurus supercluster region in the SLOW simulation compared to ROSAT}
              \label{Centaurus}
    \end{figure}
     \begin{figure}[ht]
   \centering
   \includegraphics[width=0.5\textwidth]{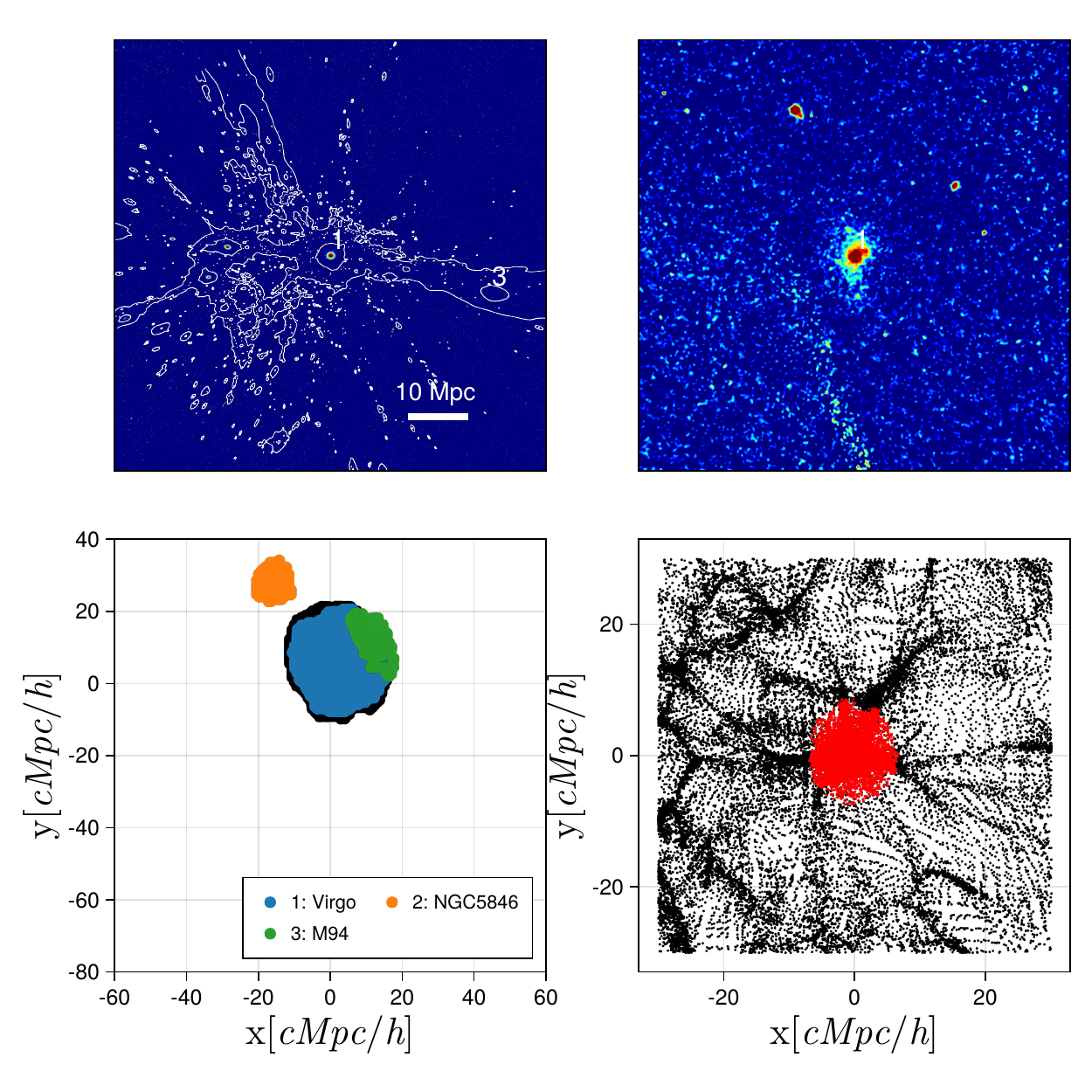}
   \caption{The Local supercluster region in the SLOW simulation compared to ROSAT}
              \label{Virgo}
    \end{figure}
    \begin{figure}[ht]
   \centering
   \includegraphics[width=0.5\textwidth]{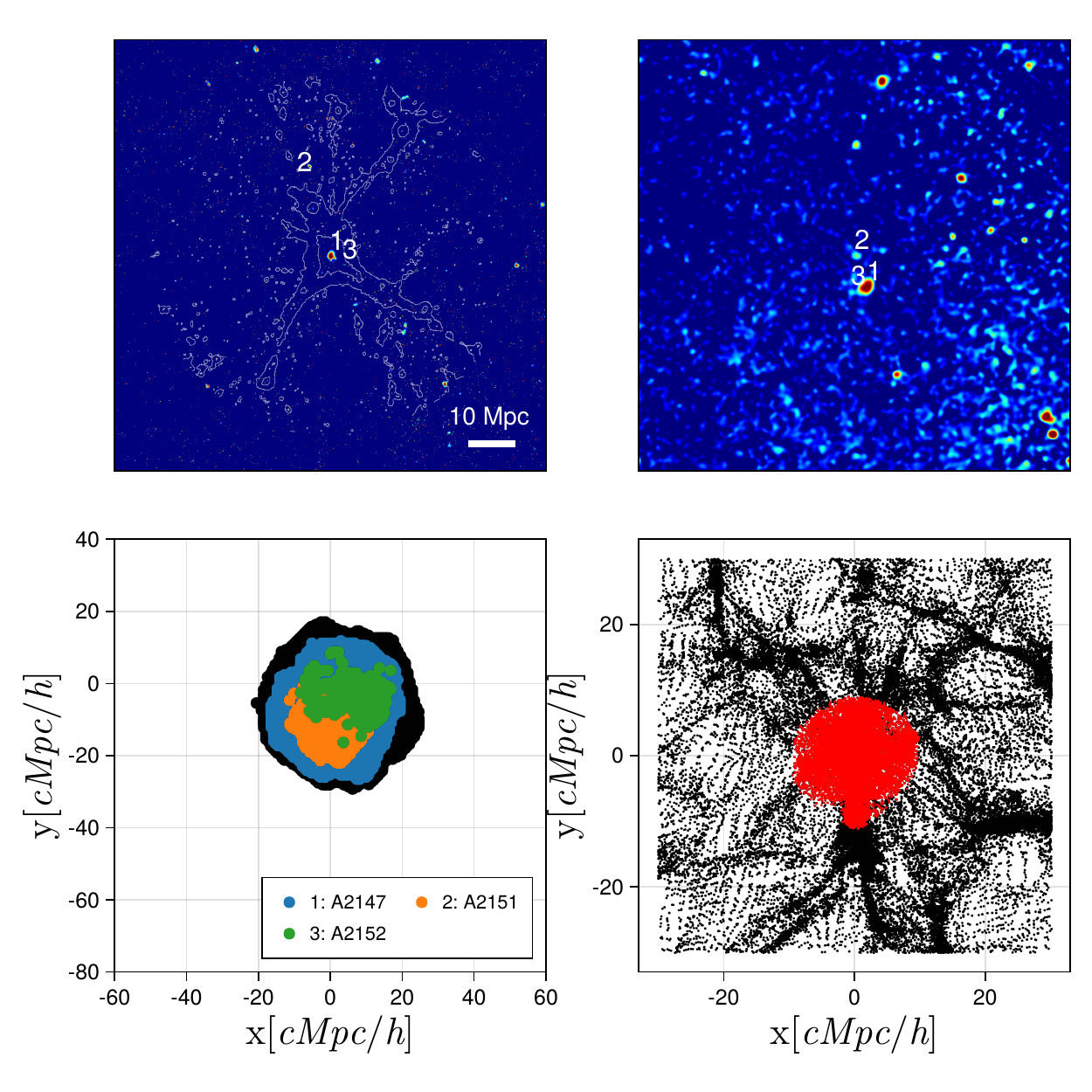}
   \caption{The Hercules supercluster region in the SLOW simulation compared to ROSAT}
              \label{Hercules}
    \end{figure}
    \begin{figure}[ht]
   \centering
   \includegraphics[width=0.49\columnwidth]{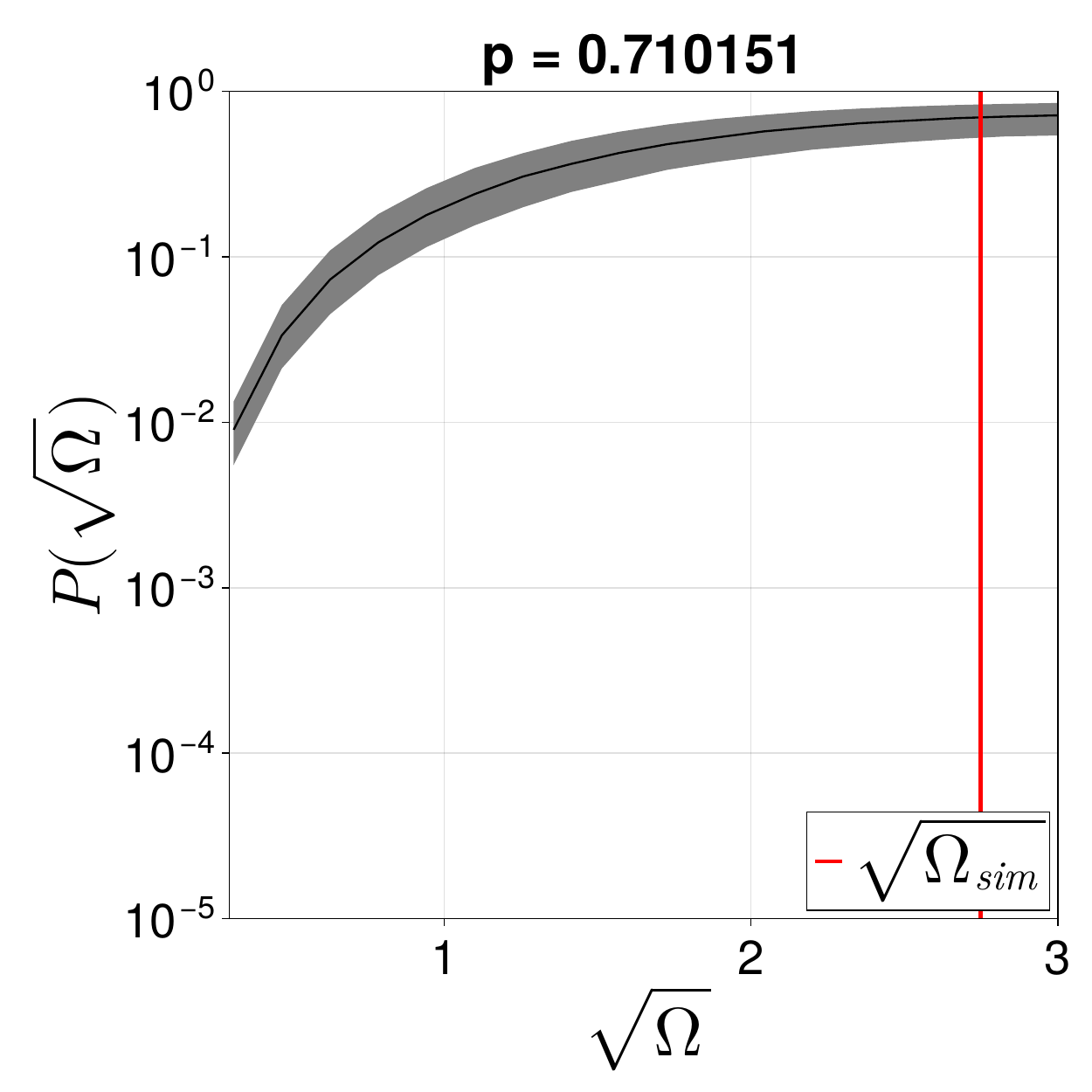}
   \includegraphics[width=0.49\columnwidth]{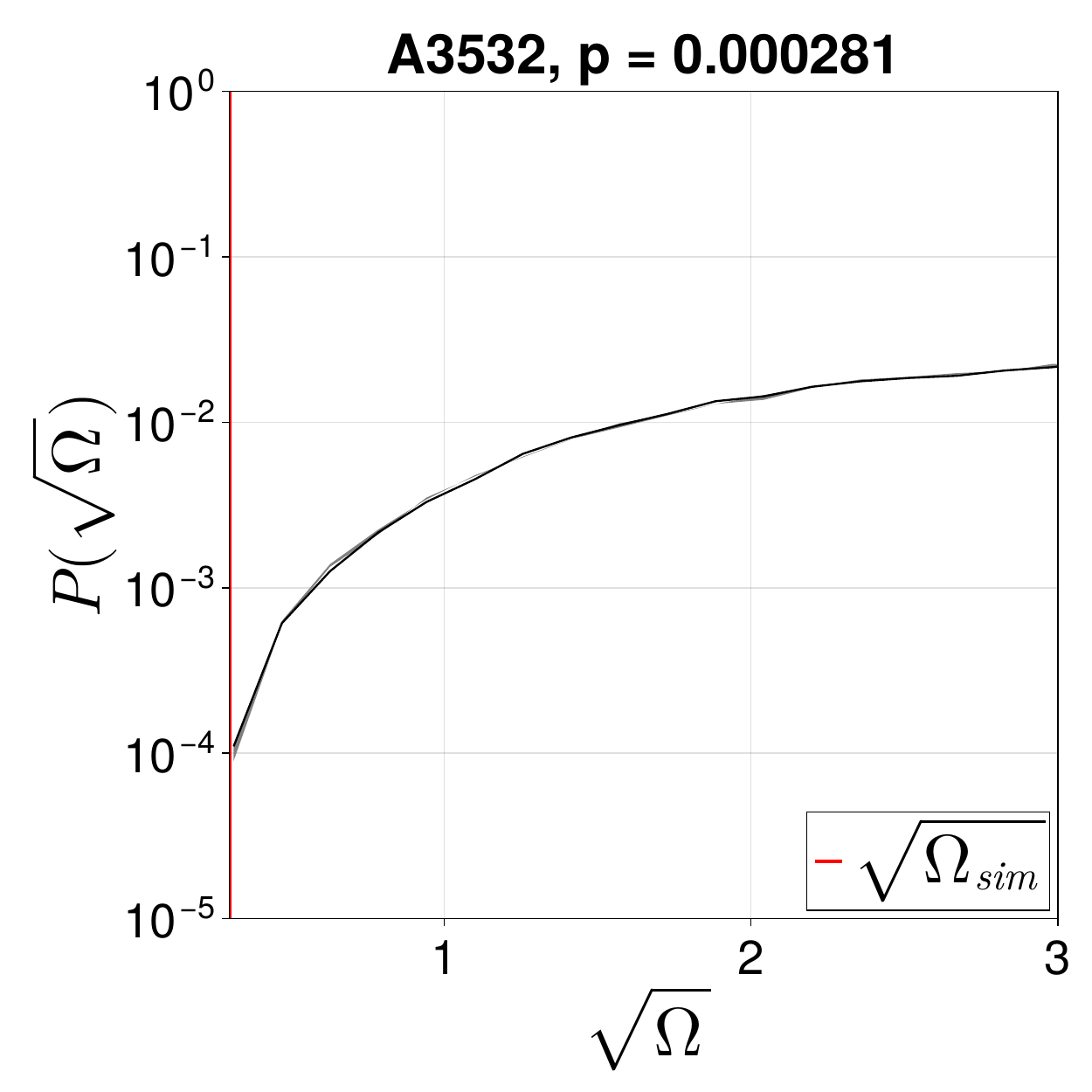}
   \caption{Two-point probabilities of the match for A3532 as identified by \citet{hernandez-martinez2024a} (left panel) with the one-point approach and in this work taking the supercluster arrangement into account (right panel).}
              \label{sig2}
    \end{figure}
    \cref{sig2} shows the how the two point significance introduced above changes between the two candidates for A3532. The better relative position of the counterpart in this work gives it a higher significance from the supercluster structure.
    \begin{figure}[ht]
   \centering
   \includegraphics[width=0.5\textwidth]{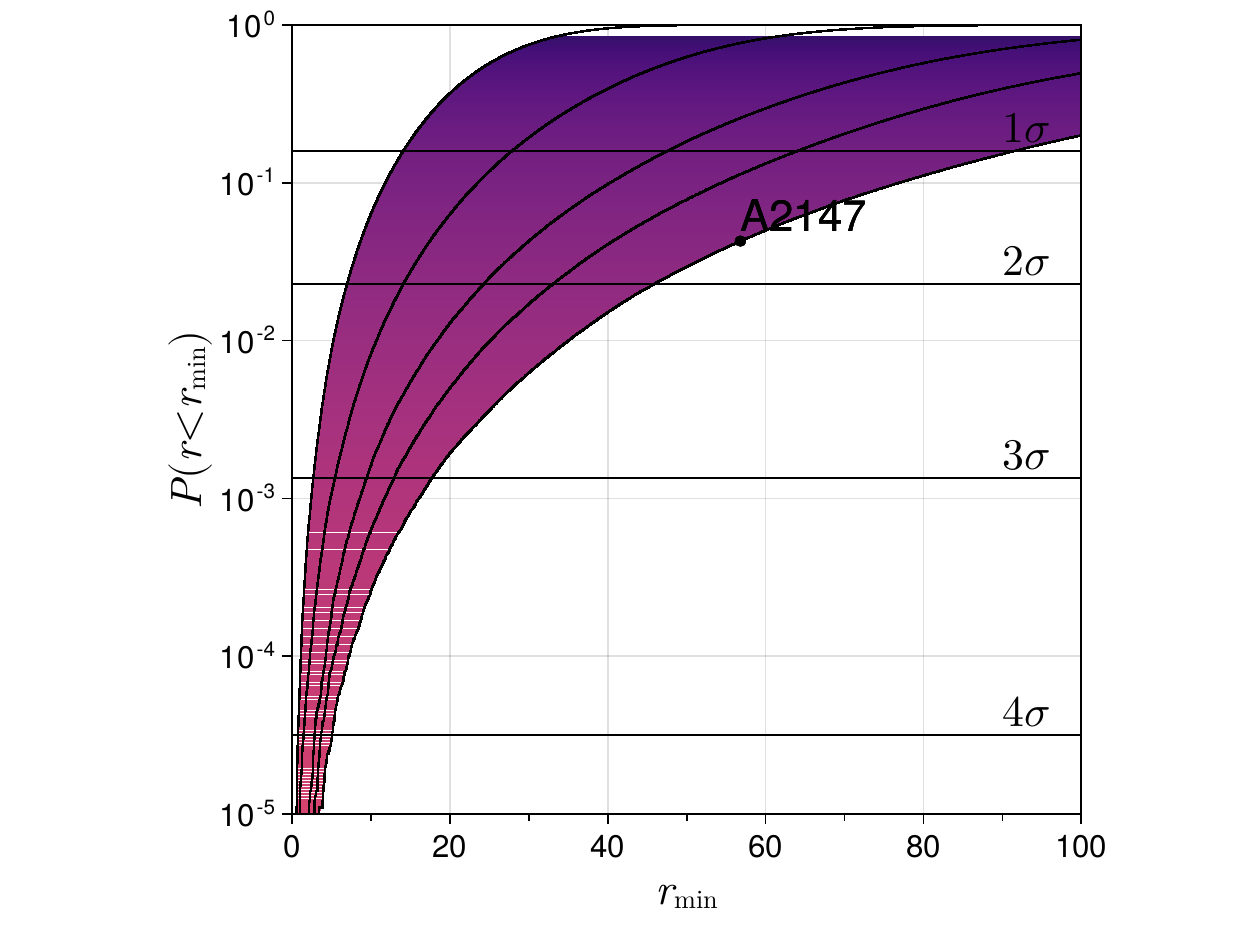}
   \caption{Match probability of A2147, following \citet{hernandez-martinez2024a}.}
              \label{A2147sig}
    \end{figure}
    Because A2147, the most massive halo of the Hercules supercluster, does not have a counterpart defined by \citet{hernandez-martinez2024a}, we evaluate the match we identify in this work with the method used in that work. The resulting probability cited in \cref{table:set} is visualized in \cref{A2147sig}.
  \section{Constraint tracer distribution}
  The SLOW simulation uses galaxy velocities from Cosmicflows-2 to build the constraints. Therefore the degree to which the simulation volume can be considered as "constrained" depends on the density of observed galaxies, as well as the accuracy of the distance measurement. To demonstrate the limitation (and how the simulation can reproduce structures even where the constraints are already sparse), we bin the number density of Cosmicflows-2 galaxies across our simulation volume and additionally weight these counts with the relative velocity error. \cref{Shapleytrace} shows this distribution in the SGX, SGY projection and demonstrates the Shapley supercluster (red contour) to be located in an area with already very low galaxy counts. The good agreement with observations in morphology and member cluster arrangement is therefore likely an indirect result of the velocity field being well-sampled along a filament connecting this structure to the more central regions where the constraint density is much higher. 
    \begin{figure}[ht]
   \centering
   \includegraphics[width=0.5\textwidth]{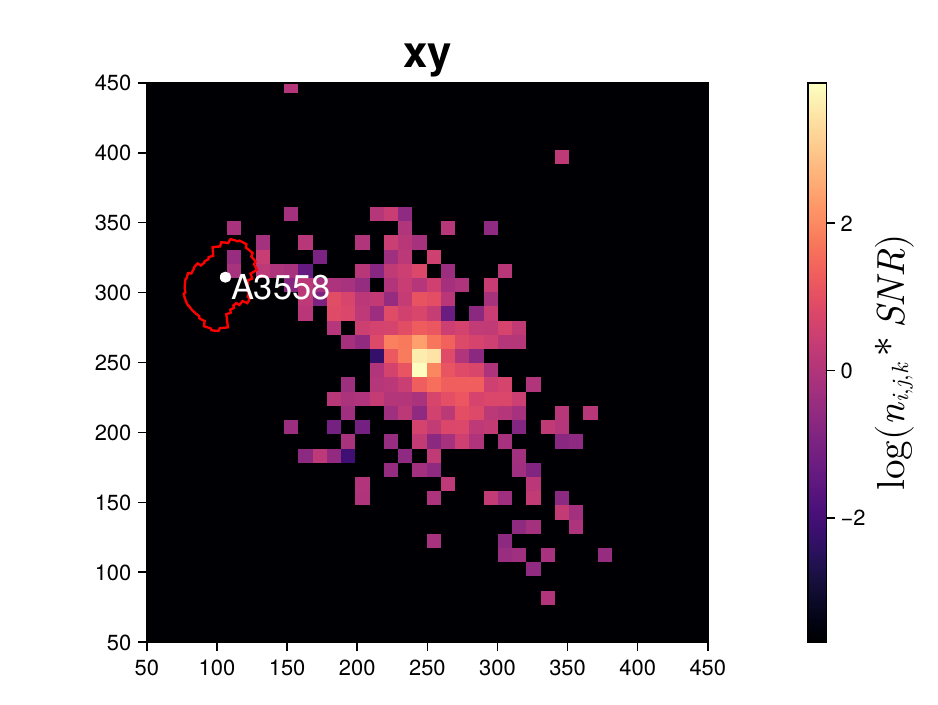}
   \caption{Weigthed counts of galaxy tracers using the signal to noise ratio (velocity amplitude:velocity error) computed from the Cosmicflows-2 database. The red contour shows the region of influence of the Shapley supercluster along with the position of A3558 in white.}
              \label{Shapleytrace}
    \end{figure}
\end{document}